\documentclass[twocolumn]{aastex63}

\usepackage[T1]{fontenc}
\usepackage{aecompl}
\usepackage{longtable}
\usepackage[para]{threeparttable}
\usepackage[normalem]{ulem}
\usepackage[utf8]{inputenc}
\usepackage{amsmath}
\usepackage{amsfonts}
\usepackage{amssymb}
\usepackage{natbib}
\usepackage{lmodern}

\newcommand{\hii}{\mbox{H~{\sc ii}~}}
\newcommand{\mgi}{\mbox{Mg~{\sc i}~}}
\newcommand{\mgii}{\mbox{Mg~{\sc ii}~}}

\newcommand{\oi}{\mbox{O~{\sc i}~}}
\newcommand{\oii}{\mbox{O~{\sc ii}~}}

\newcommand{\siii}{\mbox{Si~{\sc ii}~}}
\newcommand{\siiii}{\mbox{Si~{\sc iii}~}}

\newcommand{\ciii}{\mbox{C~{\sc iii}~}}
\newcommand{\heii}{\mbox{He~{\sc ii}~}}
\newcommand{\hei}{\mbox{He~{\sc i}~}}
\newcommand{\nai}{\mbox{Na~{\sc i}~}}
\newcommand{\cai}{\mbox{Ca~{\sc i}~}}
\newcommand{\caii}{\mbox{Ca~{\sc ii}~}}
\newcommand{\fei}{\mbox{Fe~{\sc i}~}}

\received{...}
\revised{...}
\accepted{...}

\submitjournal{AJ}

\shorttitle{Young stellar population in S242}
\shortauthors{Panja et al.}

\begin{document}

\title{Census of young stellar population in the Galactic \hii region Sh2-242}

\correspondingauthor{Alik Panja}
\email{alik.panja@gmail.com}

\author[0000-0002-4719-3706]{Alik Panja}
\affiliation{S. N. Bose National Centre for Basic Sciences, Kolkata 700106, India}

\author[0000-0003-1457-0541]{Soumen Mondal}
\affiliation{S. N. Bose National Centre for Basic Sciences, Kolkata 700106, India}

\author[0000-0002-2338-4583]{Somnath Dutta}
\affiliation{Institute of Astronomy and Astrophysics, Academia Sinica, Taipei 10617, Taiwan}

\author{Santosh Joshi}
\affiliation{Aryabhatta Research Institute of Observational Sciences, Nainital 263002, India}

\author{Sneh Lata}
\affiliation{Aryabhatta Research Institute of Observational Sciences, Nainital 263002, India}

\author{Ramkrishna Das}
\affiliation{S. N. Bose National Centre for Basic Sciences, Kolkata 700106, India}

\begin{abstract}

We present here identification and characterization of the young stellar population associated with an active star-forming site Sh2-242. We used our own new optical imaging and spectroscopic observational data, as well as several archival catalogs, e.g., Pan-STARRS 1, {\it Gaia} DR2, IPHAS, WIRCam, 2MASS, and {\it Spitzer}. Slit spectroscopic results confirm the classification of the main ionizing source BD+26\,980 as an early-type star of spectral type B0.5\,V. The spectrophotometric distance of the star is estimated as 2.08 $\pm$ 0.24 kpc, which confirms the source as a member of the cluster. An extinction map covering a large area (diameter $\sim$ 50$\arcmin$) is generated with $H$ and $K$ photometry toward the region. From the map, three distinct locations of peak extinction complexes ($A_{V}$ $\simeq$ 7--17 mag) are identified for the very first time. Using the infrared color excess, a total of 33 Class I and 137 Class II young objects are classified within the region. The IPHAS photometry reveals classification of 36 H$\alpha$ emitting sources, which might be class II objects. Among 36 H$\alpha$ emitting sources, 5 are already identified using infrared excess emission. In total, 201 young objects are classified toward S242 from this study. The membership status of the young sources is further windowed with the inclusion of parallax from the {\it Gaia} DR2 catalog. Using the optical and infrared color-magnitude diagrams, the young stellar objects are characterized with an average age of $\sim$ 1 Myr and the masses in the range 0.1--3.0 $M_\sun$. The census of the stellar content within the region is discussed using combined photometric and spectroscopic data.

\end{abstract}


\section{Introduction} \label{sec:intro}

Stellar clusters are recognized as promising astrophysical sites as their formation and early evolution take place primarily in the Galactic spiral arms \citep{lad03}. Systematic studies of young clusters probe several dominant astrophysical problems, such as the formation of stars and planetary systems to the evolution of open clusters \citep{ada06, ada10, san12}. Most of the stars originate in populous groups within giant regions of molecular clouds \citep{car00, pal02, por03}. Massive stars (>8 $M_\sun$) play an immense role to create a birthplace for next generation stars by emitting huge amounts of energy in the ultraviolet range \citep{elm77}. The Lyman continuum radiation gradually ionizes the surrounding interstellar medium (ISM) and create \hii regions. The newly formed \hii regions are the zero-age objects compared with the age of the Milky Way and are thus efficient tracers of star formation at the present epoch \citep{and14}. Zones of \hii regions are presumed to be the productive sites of second generation star formation \citep{elm98}. The early phases of stellar evolution occur within the dense regions of molecular clouds, where young stellar populations are invariably associated with significant amounts of interstellar dust and gas \citep{lad92}. Young stars associated in a cluster are thought to have formed almost simultaneously from the same progenitor molecular cloud and share similar heritage of age, distance, and chemical composition \citep{fuk00}. 
The infrared and radio wavelength surveys provide the global pictures of young star-forming regions and their formation scenarios. However, the parameters of a young cluster can be estimated from the optical/infrared photometric study of the stellar sources and the spectroscopic data of massive exciting stars associated with the cluster. Despite recent advancements in observational and theoretical prospects, the complete star formation census is still poorly understood and requires further exploration.

In this context, we present a multiwavelength survey of the Galactic \hii region Sh2-242 (S242; $\alpha_{(2000)}$ = 05$^{\rm h}$51$^{\rm m}$54$^{\rm s}$, $\delta_{(2000)}$ = +27$\degr$01$\arcmin$54$\arcsec$), located in the Taurus constellation.
 \citet{may73} first classified the main ionizing source of the region, BD+26\,980 as spectral type B0\,V. Using $UBV$ photoelectric photometry, they overestimated the distance of the star as 3.39 kpc. Using the spectroscopic observations and H$\gamma$ equivalent widths, \citet{cra74} also estimated the spectral type of BD+26\,980 as B0\,V. 
 Using the absolute magnitude from spectroscopic classification and assuming the ratio of total-to-selective absorption as 3.0, the authors calculated the distance of the star as 2.1 kpc. 
 Using $UBVRI$ photometric observations, \citet{lah87} estimated the distance to the exciting star to be 2.5 kpc and the extinction as 2.4 mag [$E(B-V)$ = 0.76 mag]. To estimate the distance, they used an absolute magnitude based on a B0\,V spectral type and assuming $A_{V}/E(B-V)$ = 3.1 or using the MK spectral types. 
 \citet{hun90} classified the spectral type of BD+26\,980 to be B0\,V using optical spectroscopy. They estimated the distance to be 2.7 kpc using simulated $B$ and $V$ photometry formed from the optical spectra of the star.

Together, these previous studies indicate that the young cluster S242 is an active site of star formation and also harbors a massive star of spectral type B0\,V. In this paper, we present a multiwavelength study on the identification and characterization of the young stellar population toward this region. 
In Section~\ref{sec:data}, we describe the observational and archival data sets used, including their reduction processes. Section~\ref{sec:res} comprises the results of spectroscopically observed bright sources and identification and classification of young stellar objects (YSOs) using archival data sets. Characterization of YSOs, such as spectral nature, age, and mass spectrum, and the cluster properties are discussed in Section~\ref{sec:dis}. The final results are summarized and concluded in Section~\ref{sec:sum}.

\begin{figure*}
\includegraphics[width=8 cm,height=7.0 cm, bb=-40 00 570 570]{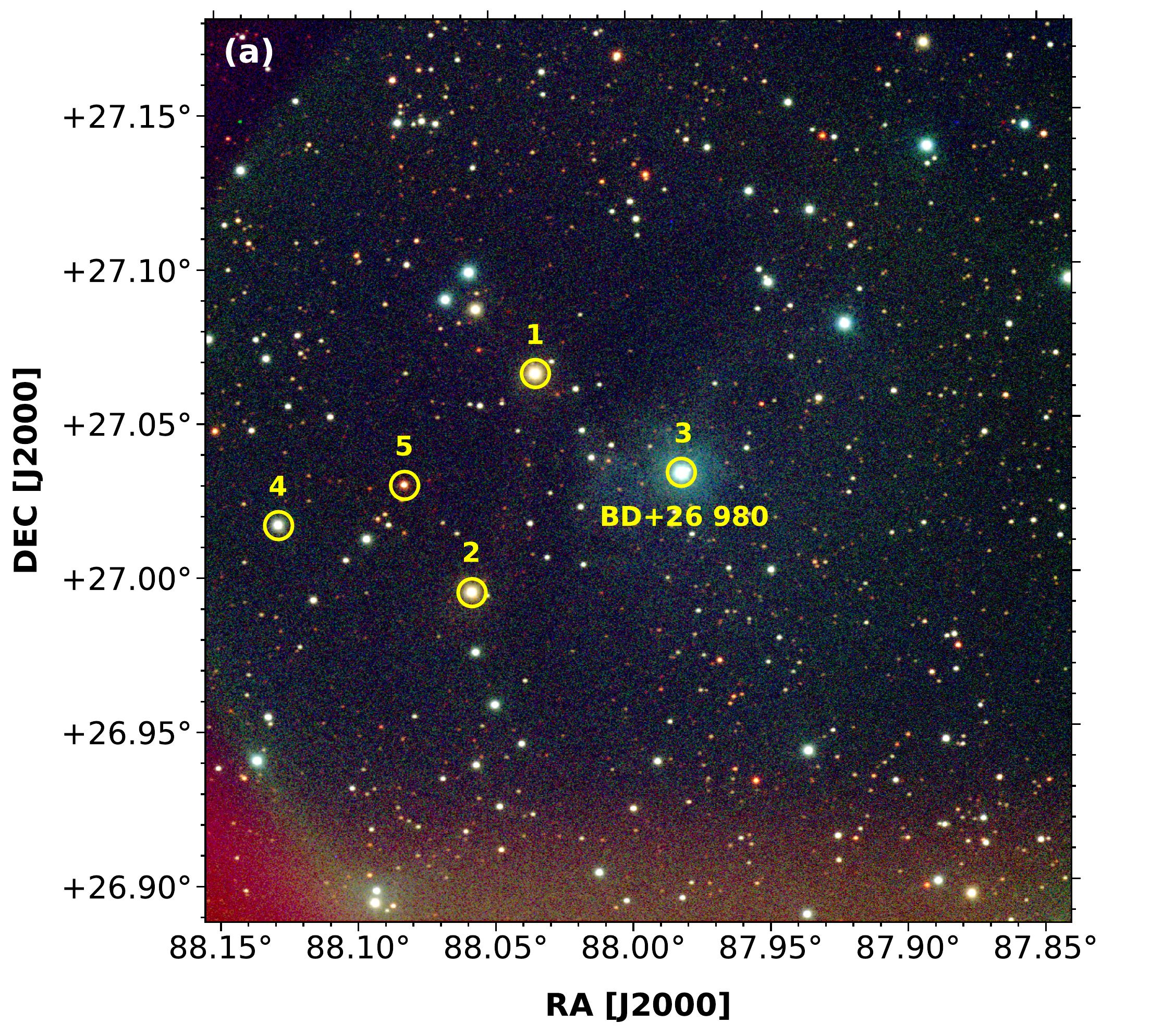}
\includegraphics[width=8 cm,height=7.0 cm, bb=-90 00 550 570]{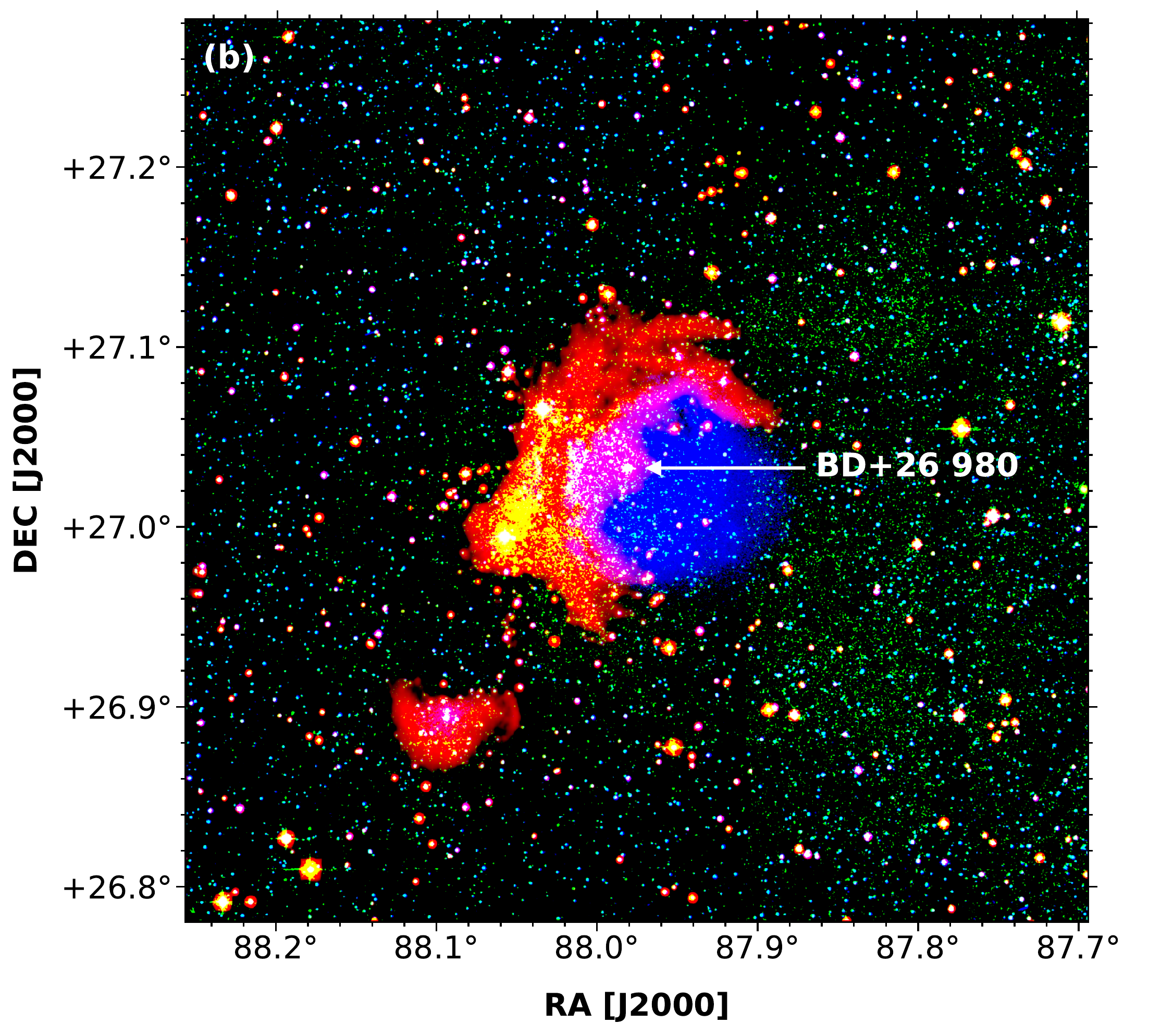}
  \caption{(a) Optical color composite image of the S242 region, created with $B$ 360 nm (blue), $V$ 550 nm (green), and $I$ 900 nm (red) bands for a sky area of 18$\times$18 arcmin$^{2}$, acquired with the 1.3~m DFOT. Spectroscopically observed sources from the 2~m HCT are numbered and marked with yellow circles and the main ionizing source BD+26\,980 is depicted. (b) Color composite image of the S242 cluster with optical and infrared counterparts taken from DSS2~$R$~0.70~$\mu$m (blue), 2MASS~K~2.2~$\mu$m (green), and {\it WISE}~$W2$~4.6~$\mu$m (red) bands.}
  \label{fig:optical_nir}
\end{figure*}

\section[]{Data Sets and Reduction}
 \label{sec:data}

\begin{table*}
\centering
  \caption{Log of Optical Photometric and Spectroscopic Observations.}
\label{tab:optical_obs}
\footnotesize
\begin{tabular}{ccccccccc}

\hline \hline

\multicolumn{1}{c}{ID} & \multicolumn{1}{c}{R.A. (J2000)} & \multicolumn{1}{c}{Decl. (J2000)} & \multicolumn{1}{c}{Date of} & \multicolumn{1}{c}{Filter/} & \multicolumn{1}{c}{Exp. time (s) $\times$} & \multicolumn{1}{c}{Airmass/} & \multicolumn{1}{c}{$V$} & \multicolumn{1}{c}{$B-V$} \\

\multicolumn{1}{c}{} & \multicolumn{1}{c}{(h:m:s)} & \multicolumn{1}{c}{(d:m:s)} & \multicolumn{1}{c}{Observations} & \multicolumn{1}{c}{Grism} & \multicolumn{1}{c}{No. of frames} & \multicolumn{1}{c}{SNR} & \multicolumn{1}{c}{(mag)} & \multicolumn{1}{c}{(mag)} \\ \hline

\multicolumn{9}{c}{Photometry} \\
\hline
Sh2-242 & 05:52:20 & +27:02:30 & 2016~Dec~19 & $B$ & 500$\times$3, 60$\times$3 & 1.018 & & \\
Sh2-242 & 05:52:20 & +27:02:30 & 2016~Dec~19 & $V$ & 500$\times$3, 60$\times$3 & 1.679 & & \\
Sh2-242 & 05:52:20 & +27:02:30 & 2016~Dec~19 & $I$ & 300$\times$3, 10$\times$3 & 1.130 & & \\
\hline

\multicolumn{9}{c}{Slit Spectroscopy} \\
\hline
 1 & 05:52:08.13 & +27:03:55.64 & 2016~Nov~03 & 7, 8 & 900$\times$1, 900$\times$1 & 18, 27 & 10.864 & 1.608 \\
 2 & 05:52:13.83 & +26:59:40.83 & 2016~Nov~03 & 7, 8 & 1200$\times$1, 1200$\times$1 & 20, 28 & 11.313 & 1.451 \\
 3 & 05:51:55.41 & +27:01:58.03 & 2016~Nov~03 & 7, 8 & 1200$\times$1, 900$\times$1 & 29, 41 & 10.128 & 0.483 \\
 4 & 05:52:30.68 & +27:01:01.15 & 2017~Jan~23 & 7, 8 & 1800$\times$1, 1800$\times$1 & 24, 26 & 11.978 & 0.914 \\
 5 & 05:52:19.65 & +27:01:46.83 & 2017~Jan~23 & 7, 8 & 1800$\times$1, 1800$\times$1 & 14, 17 & 14.246 & 1.879 \\
\hline

\multicolumn{9}{c}{Slitless Spectroscopy} \\
\hline
Sh2-242\_1 & 05:52:24.03 & +26:58:29.57 & 2017~Dec~15 & Gr5/H$_\alpha$-Br & 1800$\times$1 & 1.273 \\
Sh2-242\_2 & 05:51:44.16 & +26:58:38.72 & 2017~Dec~15 & Gr5/H$_\alpha$-Br & 1800$\times$1 & 1.361 \\
Sh2-242\_3 & 05:51:44.34 & +27:07:36.12 & 2017~Dec~15 & Gr5/H$_\alpha$-Br & 1800$\times$1 & 1.657 \\
Sh2-242\_4 & 05:52:24.45 & +27:07:31.31 & 2017~Dec~15 & Gr5/H$_\alpha$-Br & 1800$\times$1 & 1.882 \\
\hline \hline

\end{tabular}
\end{table*}

\subsection{New Observations}

\subsubsection{Optical Photometry}
 \label{sssec:opt_phot}

 Optical photometric data for $B$ (360 nm), $V$ (550 nm), and $I$ (900 nm) bands were acquired with the 1.3~m Devasthal Fast Optical Telescope (DFOT; \citealt{sag12}), Nainital, India, with different exposure times to cover bright to faint sources. The photometric observations were carried out using a 13.5 micron pixel, 2K$\times$2K Andor CCD camera, which covers a square area of about 18$\arcmin\times$18$\arcmin$ on the sky with a plate scale of 0$\farcs$53~pixel$^{-1}$. The CCD has a readout noise and gain of 7 $e^{-}$ and 2.2 $e^{-}$~ADU$^{-1}$ respectively. During the observations, the average seeing was $\sim$ 2$\arcsec$. The photometric standard field SA\,95 \citep{lan92} was observed on the same night with different airmass, to apply atmospheric and instrumental corrections to the target frames. The log of optical photometric observations is given in Table~\ref{tab:optical_obs}. The color composite image (blue: $B$, green: $V$, and red: $I$) generated from optical observations toward the S242 region is shown in the Fig.~\ref{fig:optical_nir}(a). The sources marked with yellow circles represent the stars observed with slit spectroscopy (see Section~\ref{sssec:slit_spec}) and their corresponding ID numbers (Table~\ref{tab:optical_obs}) are given. The optical/infrared three color image generated from DSS2~$R$~0.70~$\mu$m (blue), 2MASS~$K$~2.2~$\mu$m (green), and {\it WISE}~$W2$~4.6~$\mu$m (red) for the region is shown in the Fig.~\ref{fig:optical_nir}(b). The main illuminating source BD+26\,980 is depicted in both the panels.

 Optical photometric data were reduced using sophisticated tasks from the {\rm \scriptsize IRAF}\footnote{Image Reduction and Analysis Facility (IRAF) (\url{http://iraf.noao.edu/)}} software. The raw CCD images were cleaned by subtracting the median combination of bias and flat frames, following cosmic ray removal. Point sources were extracted using the {\rm \scriptsize DAOFIND} task from the {\rm \scriptsize DAOPHOT} package with necessary sky background and detection threshold limits. A point-spread function (PSF) fitting algorithm was performed on all the sources using the {\rm \scriptsize ALLSTAR} routine \citep{ste87}. Atmospheric extinction and color coefficients were calculated from observations of the standard field for corresponding filters by fitting a linear least-squares fit regression method \citep{ste92}. A set of transformation equations (Equations \ref{eqn:eq1}, \ref{eqn:eq2}, and \ref{eqn:eq3}) were used to convert the instrumental magnitudes to the standard systems as follows.

\begin{equation}
\label{eqn:eq1}
(B-V) = m_{bv} (b-v) + c_{bv}
\end{equation}
\begin{equation}
\label{eqn:eq2}
(V-I) = m_{vi} (v-i) + c_{vi} 
\end{equation}
and
\begin{equation}
\label{eqn:eq3}
V = v + m_{v} (V-I) + c_{v} 
\end{equation}

 where $B$, $V$, and $I$ are the magnitudes in standard systems and $b$, $v$, and $i$ are instrumental magnitudes corrected for the atmospheric extinction due to airmass. The details of color coefficients ($m_{bv}, m_{vi}, m_{v}$), constant terms ($c_{bv}, c_{vi}, c_{v}$), and extinction coefficients ($K_{b}, K_{v}, K_{i}$) are tabulated in Table~\ref{tab:optical_coefficients}.

\begin{table}
\centering
\caption{Color coefficients, constant terms, and extinction coefficients used for optical photometric calibrations.}
\label{tab:optical_coefficients}
\begin{tabular}{lcc}
\hline
Parameters & Constants \\\hline
Color coefficients \\
$m_{bv}$ & 1.326 $\pm$ 0.023 \\
$m_{vi}$ & 0.897 $\pm$ 0.014 \\
$m_{v}$ & $-$0.125 $\pm$ 0.011 \\
Zero-point constants \\
$c_{bv}$ & $-$1.019 $\pm$ 0.027 \\
$c_{vi}$ & 0.384 $\pm$ 0.016 \\
$c_{v}$ & $-$2.024 $\pm$ 0.013 \\
Extinction coefficients \\
$K_{b}$ & 0.262 $\pm$ 0.010 \\
$K_{v}$ & 0.229 $\pm$ 0.005 \\
$K_{i}$ & 0.103 $\pm$ 0.005 \\
\hline
 \end{tabular}
 \end{table}

\begin{figure}
  \includegraphics[width=10 cm,height=8 cm, bb=20 00 370 250]{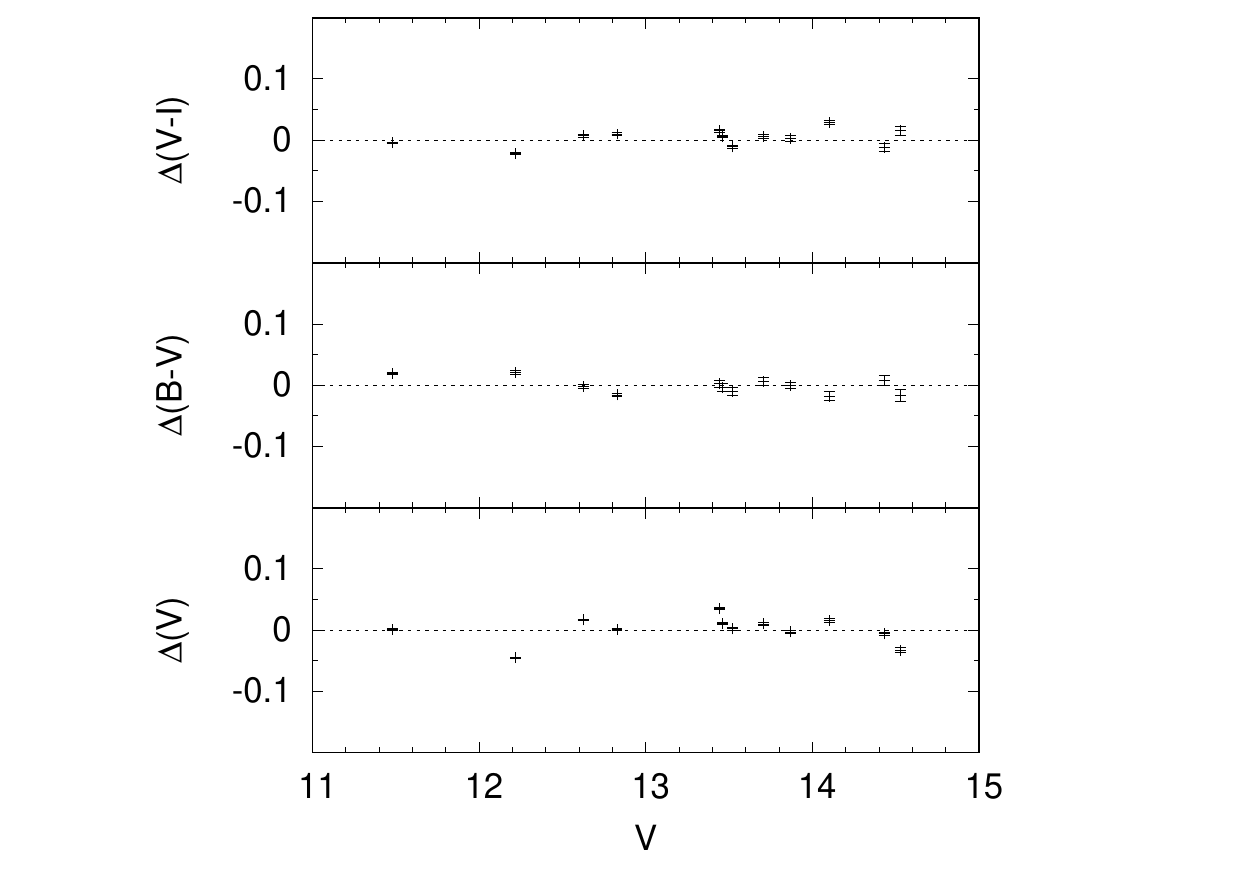}
  \caption{Plot of residuals between transformed colors and magnitudes with standard values as a function of standard $V$ magnitudes.}
  \label{fig:phot_standard}
\end{figure}

 A total of 12 standard stars was used to calculate the color coefficients and zero-point constants in the standard field. Fig.~\ref{fig:phot_standard} shows the variation of residuals between transformed and standard colors and magnitudes as a function of $V$ magnitude.

 The world coordinate system on the physical frames was obtained by selecting 24 unsaturated, isolated, and moderately bright stars from the 2MASS catalog for the same field. We used the {\rm \scriptsize IRAF} tasks {\rm \scriptsize CCMAP}, and {\rm \scriptsize CCSETWCS} and the WCSTOOLS package to obtain the astrometric solutions.

\subsubsection{Optical Slit Spectroscopy}
 \label{sssec:slit_spec}

 Optical slit spectroscopic observations for five bright sources within the region were carried out using the Himalaya Faint Object Spectrograph and Camera (HFOSC) on the 2~m Himalayan Chandra Telescope (HCT). Spectra were taken with Grism 7 (380 - 684 nm) and Grism 8 (580 - 830 nm) with resolutions of 1330 and 2190, respectively. The spectroscopic standard star Feige\,34 \citep{oke90} was observed on the same night for flux calibration. The FeAr and FeNe arc lamp observations were conducted immediately after the target observations for wavelength calibration.  The log of optical spectroscopic observations is shown in Table~\ref{tab:optical_obs}.

 After being rectified by bias subtraction and cosmic ray correction, the monodimensional spectra were extracted using the {\rm \scriptsize APALL} task in the {\rm \scriptsize IRAF} software. The spectra were then wavelength calibrated from corresponding lamp observations. The data were corrected for atmospheric extinction and instrument sensitivity availing the standard star observations.

\subsubsection{Optical Slitless Spectroscopy}

 Slitless spectroscopic data were obtained from the HCT to detect H$\alpha$ emission line stars toward the region. The data were acquired using a broadband H$\alpha$ filter (630 - 674 nm) in combination with Grism 5 (520 - 1030 nm) in slitless mode (Table~\ref{tab:optical_obs}). The 2K$\times$2K CCD has a field of view (FOV) of 10$\times$10 arcmin$^{2}$ with an image scale of 0$\farcs$296~pixel$^{-1}$. The resolution of Grism 5 is 870. The data were taken for four overlapping frames covering a total sky area (18$\arcmin\times$18$\arcmin$) similar to that of the optical photometric observations (Section~\ref{sssec:opt_phot}). The H$\alpha$ emitting sources show certain enhancements in their spectra over the continuum. We have visually identified three sources having prominent H$\alpha$ emissions from these observations.

\subsection{Optical/NIR Photometry from Pan-STARRS 1}

As a complementary data set to the observed optical photometry (Section~\ref{sssec:opt_phot}), we used optical/NIR imaging from the Panoramic Survey Telescope and Rapid Response System 1 (Pan-STARRS 1, or PS1; \citealt{cha19}). The PS1 camera consists of a mosaic of 60 edge-abutted 4846$\times$4868 pixel detectors, with 10~$\mu$m pixels subtending 0$\farcs$258. The PS1 provides homogeneous and deeper coverage photometry in five broad passbands {\large g\textsubscript{P1}, r\textsubscript{P1}, i\textsubscript{P1}, z\textsubscript{P1}}, and {\large y\textsubscript{P1}}, ranging from 400 nm to 1000 nm \citep{stu10}. The effective wavelengths for the five filters are roughly 480, 620, 750, 870, and 960 nm, respectively, similar to those used by the SDSS \citep{yor00}, with most significant difference being the replacement of the Sloan u band with a NIR band {\large y\textsubscript{P1}}. The PS1 image processing, astrometry, and photometry are detailed in \citet{mag16a, mag16b} and the relative and absolute calibration survey are reported in \citet{sch12} and \citet{ton12}. The typical single-epoch 5$\sigma$ photometric depths in the corresponding five passbands are 22.0, 21.8, 21.5, 20.9, and 19.7 mag (AB), respectively \citep{cha19}. The empirical components of the adopted extinction vector are taken from \citet{sch16} and given in Table~\ref{tab:ps1_ext}.

\begin{table}
\centering
\caption{Extinction vector, {\it R}, adopted in this work for the Pan-STARRS 1 catalog, are based on \citet{sch16}.}
\label{tab:ps1_ext}
\begin{tabular}{ccccc}
\hline \hline
{\large g\textsubscript{P1}} & {\large r\textsubscript{P1}} & {\large i\textsubscript{P1}} & {\large z\textsubscript{P1}} & {\large y\textsubscript{P1}} \\\hline \hline
3.384 & 2.483 & 1.838 & 1.414 & 1.126 \\
\hline
\end{tabular}
\end{table}

\subsection{Archival Data from {\it Gaia} DR2}

The {\it Gaia} Data Release 2 ({\it Gaia} DR2; \citealt{gai18}) provides five parameter (position, proper motion, and parallax) astrometric results for over 1.3 billion sources from the observations of the European Space Agency {\it Gaia} satellite \citep{gai16}. The parallaxes of the sources toward the S242 cluster are collected from the {\it Gaia} DR2 archive\footnote{\url{https://gea.esac.esa.int/archive/}}. The distance and membership status of the sources are estimated by restricting the data to only positive parallaxes. The positive parallaxes with relative uncertainties typically below 20\% are primarily considered as reliable astrometry \citep{lur18}.

\subsection{Archival Data from IPHAS DR2}

The INT/WFC Photometric H$\alpha$ Survey of the Northern Galactic Plane (IPHAS; \citealt{dre05}) is an imaging survey covering an 1800 deg$^2$ sky in broadband Sloan $r$ (624 nm) and $i$ (774.3 nm), and narrowband H$\alpha$ (656.8 nm) filters using the Wide Field Camera (WFC) on the 2.5~m Isaac Newton Telescope (INT) in La Palma. The WFC generates a mosaic of four CCDs at a pixel scale of 0$\farcs$33~pixel$^{-1}$. This survey offers an unique facility to detect H$\alpha$ emission line objects by comprehensive CCD photometry of point sources at visible wavelengths. The photometric data for $r$, $i$, and H$\alpha$ bands were obtained from the IPHAS DR2 catalog \citep{bar14} for the S242 cluster.

\begin{table*}
\renewcommand{\tabcolsep}{2.0pt}
\begin{center}
\caption{Details of the spectroscopically observed stars.}
\tiny
\label{tab:spectro}
\begin{tabular}{crrcrrrrcccc}
\hline

\multicolumn{1}{c}{Star} & \multicolumn{1}{c}{R.A. (J2000)} & \multicolumn{1}{c}{Decl. (J2000)} & \multicolumn{1}{c}{Spectral} & \multicolumn{1}{c}{$J$} & \multicolumn{1}{c}{$H$} & \multicolumn{1}{c}{$K$} & \multicolumn{1}{c}{Spectroscopic} & \multicolumn{1}{c}{Distance} & \multicolumn{1}{c}{Distance} & \multicolumn{1}{c}{Distance from} & \multicolumn{1}{c}{Remarks} \\

\multicolumn{1}{c}{ID} & \multicolumn{1}{c}{(h:m:s)} & \multicolumn{1}{c}{(d:m:s)} & \multicolumn{1}{c}{Type} & \multicolumn{1}{c}{(mag)} & \multicolumn{1}{c}{(mag)} & \multicolumn{1}{c}{(mag)} & \multicolumn{1}{c}{$A_{V}$ (mag)} & \multicolumn{1}{c}{Modulus} & \multicolumn{1}{c}{(pc)} & \multicolumn{1}{c}{{\it Gaia} (pc)} & \multicolumn{1}{c}{} \\

\multicolumn{1}{c}{(1)} & \multicolumn{1}{c}{(2)} & \multicolumn{1}{c}{(3)} & \multicolumn{1}{c}{(4)} & \multicolumn{1}{c}{(5)} & \multicolumn{1}{c}{(6)} & \multicolumn{1}{c}{(7)} & \multicolumn{1}{c}{(8)} & \multicolumn{1}{c}{(9)} & \multicolumn{1}{c}{(10)} & \multicolumn{1}{c}{(11)} & \multicolumn{1}{c}{} \\

\hline
1 & 05:52:08.13 & +27:03:55.64 & K1\,V-III & 7.977 $\pm$ 0.009 & 7.309 $\pm$ 0.057 & 7.065 $\pm$ 0.009 & 2.38 $\pm$ 0.06 & $...$ & $...$ & 1141 $\pm$ 64 & Foreground \\

2 & 05:52:13.83 & +26:59:40.83 & K0\,V-III & 8.709 $\pm$ 0.015 & 8.088 $\pm$ 0.019 & 7.884 $\pm$ 0.023 & 2.12 $\pm$ 0.02 & $...$ & $...$ & 1103 $\pm$ 70 & Foreground \\

3 & 05:51:55.41 & +27:01:58.03 & B0.5\,V & 9.155 $\pm$ 0.021 & 9.110 $\pm$ 0.025 & 8.982 $\pm$ 0.020 & 1.80 $\pm$ 0.04 & 11.58 $\pm$ 0.05 & 2076 $\pm$ 239 & 2079 $\pm$ 192 & Member \\

4 & 05:52:30.68 & +27:01:01.15 & G9\,V & 10.399 $\pm$ 0.019 & 10.007 $\pm$ 0.024 & 9.868 $\pm$ 0.019 & 0.24 $\pm$ 0.03 & 6.19 $\pm$ 0.04 & 173 $\pm$ 16 & 769 $\pm$ 32 & Foreground \\

5 & 05:52:19.65 & +27:01:46.83 & G0\,V & 10.006 $\pm$ 0.019 & 9.383 $\pm$ 0.024 & 9.004 $\pm$ 0.017 & 3.28 $\pm$ 0.03 & $...$ & $...$ & 2764 $\pm$ 506 & Background \\

\hline
\end{tabular}
\end{center}
\begin{tablenotes}
\tiny
 \item {\bf Notes:} \\
 \item (1) ID Number of the spectroscopically observed stars. \\
 \item (2-3) Equatorial coordinates of the stars in degrees. \\
 \item (4) Spectral types estimated from the spectroscopic observations. \\
 \item (5-7) Photometric parameters of the stars from 2MASS catalog \citep{skr06}. \\
 \item (8-9) Visual extinction ($A_{V}$) and distance modulus are calculated from spectral types and infrared photometry. \\
 \item (10) Spectroscopic distances of the stars. \\
 \item (11) Distance of the stars from {\it Gaia} DR2 catalog \citep{gai18}. \\
\end{tablenotes}
\end{table*}

\subsection{Infrared Archival Data}

\subsubsection{Near-Infrared Data: WIRCam and 2MASS}

The Wide-field InfraRed Camera (WIRCam; \citealt{pug04}) is the NIR mosaic imager mounted at the prime focus of the 3.6~m Canada-France-Hawaii Telescope (CFHT) on Maunakea, Hawaii. The WIRCam consists of four 2048$\times$2048 HAWAII2-RG detectors covering a field of view of 20 arcmin$^2$ with a pixel scale of 0$\farcs$3. The deep NIR images in $J$ (1.25 $\mu$m), $H$ (1.63 $\mu$m), and $K$ (2.14 $\mu$m) bands\footnote{\url{http://www.cfht.hawaii.edu/Instruments/Filters/wircam.html}} for the S242 region were collected from the CFHT archive\footnote{\url{http://www.cadc-ccda.hia-iha.nrc-cnrc.gc.ca/en/cfht/}} for Proposal ID 06BF14 and 06BF96, respectively. The observations were carried out under principal investigator Lise Deharveng on December 28, 2006 and January 01, 2007 accordingly. The raw data were optimized using the Interactive Data Language (IDL) based interface the SIMPLE Imaging and Mosaicking Pipeline (SIMPLE; \citealt{wan10}). The astrometric and photometric reductions were performed in a similar way, as are outlined in \citet{dut18}. Briefly the dithered images were mosaicked using median combination technique. The astrometric calibrations of the combined images were performed in comparison with the 2MASS reference frames. The point sources in the reduced astrometric frames were identified by using PSF fitting algorithm from the {\rm \scriptsize DAOFIND} package \citep{ste92} in the {\rm \scriptsize IRAF} software. Photometric calibrations of the WIRCam frames were performed in comparison with the 2MASS catalog considering all the sources with magnitude uncertainty $<$ 0.1 mag.

The 2MASS Point Source Catalog (PSC; \citealt{skr06}) photometry in $J$, $H$, and $K$ bands were taken as a complementary data set. In an attempt to avoid the inclusion of saturated sources in WIRCam photometry, we replaced all the sources in WIRCam with 2MASS magnitudes for $J$ < 13 mag, $H$ < 12.5 mag, and $K$ < 12 mag \citep{dut18}. A photometric uncertainty $<$ 0.1 mag for all the bands was considered as quality criteria for reliable photometry, which provides S/N $\geq$ 10.

\subsubsection{Mid-Infrared Data: {\it Spitzer}}

We obtained MIR photometry for point sources toward the S242 region from the {\it Spitzer} Warm Mission \citep{hor12} survey. Magnitudes from the Infrared Array Camera (IRAC; \citealt{faz04}) [3.6] and [4.5] $\mu$m bands with a pixel scale of 1$\farcs$2~pixel$^{-1}$ were downloaded from the highly reliable Glimpse360\footnote{\url{http://www.astro.wisc.edu/sirtf/glimpse360/}} catalog (Program Id: 61070, PI: Whitney, Barbara A). We restricted the sources with photometric uncertainty $\sigma$ < 0.2 mag for all the IRAC bands to achieve good quality photometric catalog.

\subsection{Multiwavelength Catalog}

The final catalog was generated by matching different optical to infrared data sets in stages. The $BVI$ catalog was built by cross-matching all the sources detected from optical photometry (Section~\ref{sssec:opt_phot}) with a radial tolerance of 2$\arcsec$. As the seeing ($\sim$ 2$\arcsec$) was not sufficiently good, we used a matching radius of 2$\arcsec$ to match the sources detected in optical photometry. We performed several test matches by increasing the radial distance from 1$\arcsec$ to 3$\arcsec$ in steps of 0$\farcs$1 to pick up the suitable matching radius for each catalog data. In case of multiple sources matched within a given matching radius, we have taken the nearest one as the preferred match. To match the sources detected from WIRCam $JHK$ bands, a matching radius of 1$\arcsec$ was used. $JHK$, IRAC, and IPHAS catalogs were matched within a matching radius of 1$\arcsec$. We have adopted the matching radius of 1$\arcsec$ to match the sources selected from optical and infrared catalog data.

We have used histogram turn over method to estimate the completeness of our utilized data sets. In general, the completeness limits are calculated from histograms, where the logarithmic distribution of the sources deviates from the linear distribution. The completeness limits for different data sets for the S242 region are $V$ = 17.8 mag, $J$ = 18.4 mag, $H$ = 17.6 mag, $K$ = 17.2 mag, [3.6] = 16.8 mag, and [4.5] = 16.4 mag, respectively. However, there may be several additional factors, such as saturation caused by bright luminous sources, variable reddening, stellar crowding, telescope detection sensitivity, etc., that can constrain the completeness of different data sets \citep{jos13}.

\section{Results}
 \label{sec:res}

\subsection{Spectral Classification of the Spectroscopically Observed Stars}

\begin{figure*}
\centering
\includegraphics[width=8.5 cm, height=7.0 cm]{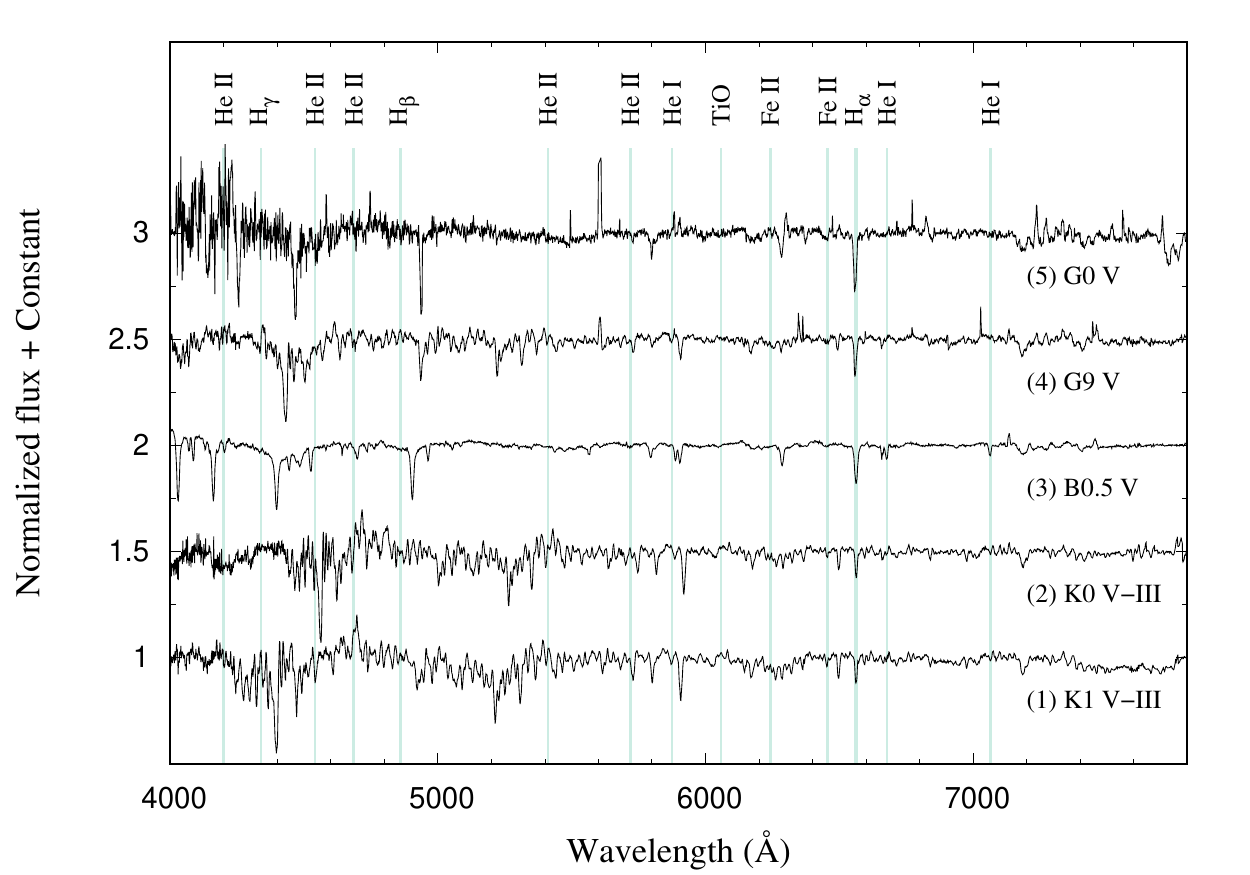}
\includegraphics[width=8.5 cm, height=7.0 cm]{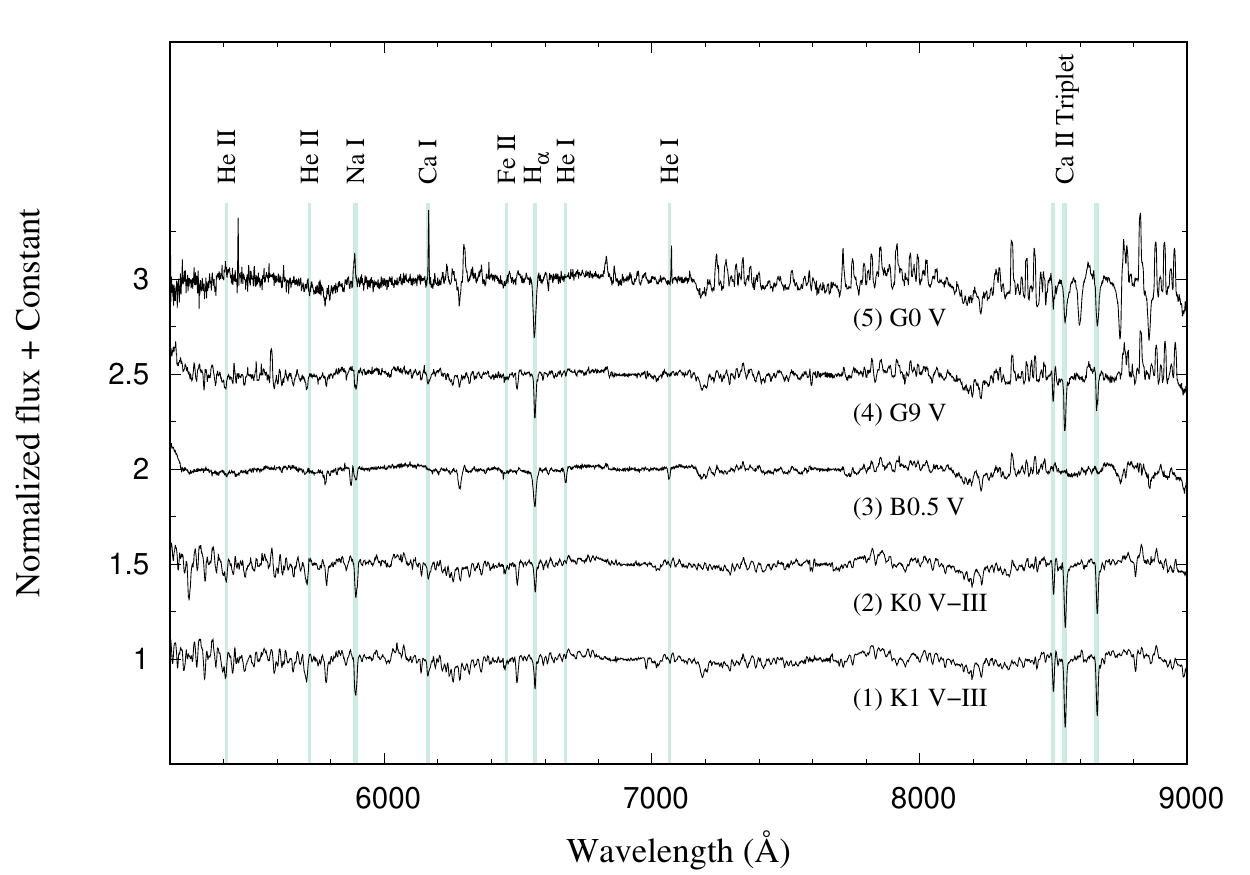}
  \caption{Flux-calibrated normalized spectra for the spectroscopically observed stars towards S242, obtained with the 2-m HCT using Grism 7 (left) and Grism 8 (right) respectively. The important emission and absorption line features are marked.}
   \label{fig:spectra7}
\end{figure*}

 The target stars for spectroscopic observations were selected based on their brightness ($J$ < 11 mag) within the cluster region. The details of observations for each star are listed in Table~\ref{tab:optical_obs}. The flux-calibrated normalized spectra of the observed stars obtained with Grism 7 and 8 are presented in Fig.~\ref{fig:spectra7}.

 Spectral classifications were done by comparison to the spectral indices of \citet{dan94}, \citet{kob12}, and \citet{her04}, and comparison with the spectral atlas of \citet{jac84}, \citet{wal90}, and \citet{tor93}. The classification scheme relied on the marking and utilizing of strong conspicuous features for any spectral range. For early-type stars (B, A, and F), we compared the strength of atomic absorption lines, such as hydrogen Balmer series (H$\delta$ $\lambda$4102 \r {A}, H$\gamma$ $\lambda$4340 \r {A}, H$\beta$ $\lambda$4861 \r {A}, H$\alpha$ $\lambda$6563 \r {A}), \hei ($\lambda\lambda$5876, 6678, 7065 \r {A}), and \heii ($\lambda\lambda$4200, 4541, 4686, 5411, 5720 \r {A}) lines. Whereas for cooler stars like G-type or later than that, different metallic line features such as \mgi triplet ($\lambda\lambda$5167, 5172, 5183 \r {A}), \mgii ($\lambda\lambda$4481, 6347 \r {A}), \cai ($\lambda\lambda$6122, 6162 \r {A}), and \fei ($\lambda\lambda$6495, 7749, 7834 \r {A}) are used. We also adopted certain constraints on the specific line features. The absence of \heii $\lambda$5411 \r {A} line in any spectra limits the spectral type to B0.5 or later \citep{kob12}, while the absence of \hei $\lambda$5876 \r {A} constrains the spectral type to later than A0 \citep{lun14}. \siiii $\lambda$4552 \r {A} and \oii $\lambda$4650 \r {A} show maximum strength at spectral type B0.5\,V, and \heii $\lambda$4686 \r {A} is last seen up to types B0.5$-$B0.7\,V \citep{wal90}. The declining strength of \ciii $\lambda$4070 \r {A} and \oii $\lambda$4650 \r {A} blends are used as an additional criteria to classify the stars in the spectral range B1$-$B2\,V. \hei line strength is maximum for B2 type stars, whereas for later than that \siii $\lambda\lambda$4128$-$4130 \r {A} and \mgii $\lambda$4481 \r {A} increase distinctly \citep{wal90}. \hei $\lambda$6678 \r {A} appears strongest at O9\,V, before disappearing at B8\,V \citep{dan94}. The weak presence of \oi $\lambda$7776 \r {A} is notable at B2\,V, strengthens to maximum at A5\,V and disappears at G0\,V. The appearance of \fei $\lambda$6495 \r {A} is evident at A2\,V and grows in strength to K0\,V \citep{dan94}. The presence of \fei $\lambda\lambda$7749, 7937 \r {A} in any spectrum is an indication of K dwarfs \citep{all95}. We also compared the equivalent widths of \heii $\lambda$5411 \r {A}, \hei $\lambda$5876 \r {A} and H$\alpha$ $\lambda$6563 \r {A} for B-type stars, and \nai ($\lambda\lambda$5890$-$5896 \r {A}), H$\alpha$ and \caii triplet ($\lambda\lambda$8498, 8542, 8662 \r {A}) for later-type stars with the spectral indices of \citet{dan94}, \citet{kob12}, and \citet{lun14}.

The spectroscopic analysis of the observed sources produced only one massive and early-type star (B0.5\,V). Other spectroscopically observed sources were found to be late-type (either G or K) stars. The star ID 1 shows weaker H$\alpha$ $\lambda$6563 \r {A} and \nai $\lambda\lambda$5893, 8195 \r {A} absorption features, presence of \fei $\lambda\lambda$6495, 7747, 7834 \r {A} and \caii triplet $\lambda\lambda$8498, 8542, 8662 \r {A} indicative of an early-K type star (K1\,V-III). The star ID 2 is classified as K0\,V-III, from similar diagnostic as ID 1. The spectrum of star ID 3 shows strong \hei $\lambda$5876 \r {A} and H$\alpha$ absorption feature along with the presence of \heii $\lambda$4200 \r {A} and \hei $\lambda\lambda$6678, 7065 \r {A} lines. As no signature of \heii $\lambda$5411 \r {A} was detected from the spectrum, we categorized the star as B0.5\,V spectral type. 
The stars ID 4 and 5 were classified as G9\,V and G0\,V type, respectively, for showing weak and narrow H$\alpha$ absorption features and presence of \mgi triplet at $\lambda\lambda$5167, 5172, 5183 \r {A}. We designated the stars with luminosity class V/III, as their spectral resemblance with the main-sequence/giants are better than super-giants. Also in some cases it was difficult to properly distinguish between the main-sequence and the giant stars. Based on the low-resolution spectroscopy, an uncertainty of $\pm$1 spectral subtype for early-type stars up to F- and $\pm$3 subtype for G-type and later stars is expected. The photometric and spectroscopic details of the five observed stars are tabulated in Table~\ref{tab:spectro}.

\subsection{Reddening and Membership of the Spectroscopically Observed Stars}

 Membership estimation of the observed stars toward a cluster region is a crucial step to quantify the essential cluster parameters. The relevant parameters used to ascertain the spectrophotometric distances for each star are listed in Table~\ref{tab:spectro}. We used the spectral types and infrared photometry ($J$, $H$, and $K$) to determine the distances of the spectroscopically observed bright sources \citep{dut15}. We have estimated the spectroscopic $A_{V}$ of individual sources, using the relation $E(J-H) = (J-H) - (J-H)_{0}$, and similar relations simultaneously for other two bands, where $(J-H)$ is the observed color and $(J-H)_{0}$ being the intrinsic color. The intrinsic distance modulus ($J_{0}-M_{J}$), ($H_{0}-M_{H}$), and ($K_{0}-M_{K}$) were calculated from reddening $A_{V}$, absolute ($M_{J}, M_{H}, M_{K}$) and observed ($J, H, K$) magnitudes. The intrinsic values of magnitudes and colors were taken from \citet{pec13}. The intrinsic distance modulus of star ID 3 was calculated as 11.58 $\pm$ 0.05 mag, which corresponds to a distance of 2.08 $\pm$ 0.24 kpc. Additionally from the {\it Gaia} DR2 catalog, the distance of the source is derived as 2.08 $\pm$ 0.19 kpc, which is in close agreement with our estimated spectrophotometric distance. Out of the five spectroscopically observed stars, we assigned BD+26\,980 (star ID 3) of spectral type B0.5\,V, a massive member of the cluster. Others are either foreground or background stars as mentioned in Table~\ref{tab:spectro}.

\begin{figure}
  \includegraphics[width=10 cm,height=7 cm]{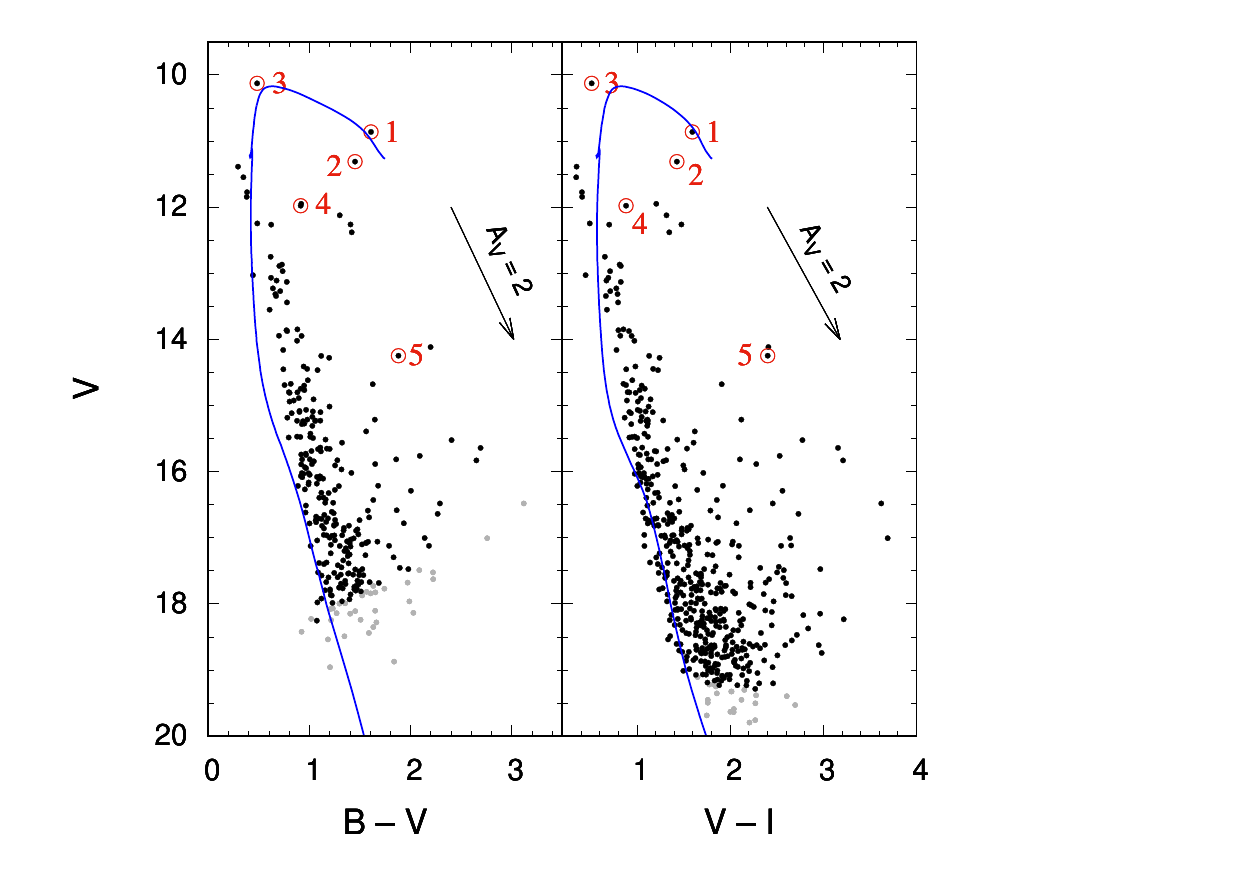}
  \caption{Distribution of sources from optical photometry in the CM diagrams. Spectroscopically observed sources are numbered and marked with red circles. The ZAMS from \citet{gir02}, corrected for the cluster distance of 2.08 kpc and reddening $E(B-V)$ = 0.56 mag and $E(V-I)$ = 0.70 mag (Table~\ref{tab:spectro}), is shown with blue solid lines.}
  \label{fig:opt_cmd}
\end{figure}

\renewcommand{\tabcolsep}{2.0pt}
\begin{table*}
\begin{center}
\caption{Photometric catalog of optically observed point sources toward the S242 region. The complete catalog is available in the electronic version.}
\tiny
\label{tab:bvi_phot}
\begin{tabular}{cccrrrrrrrr}
\hline

\multicolumn{1}{c}{Sl.} & \multicolumn{1}{c}{R.A. (J2000)} & \multicolumn{1}{c}{Decl. (J2000)} & \multicolumn{1}{c}{$V$} & \multicolumn{1}{c}{$B-V$} & \multicolumn{1}{c}{$V-I$} & \multicolumn{1}{c}{$J$} & \multicolumn{1}{c}{$H$} & \multicolumn{1}{c}{$K$} & \multicolumn{1}{c}{[3.6] $\mu$m} & \multicolumn{1}{c}{[4.5] $\mu$m} \\

\multicolumn{1}{c}{No.} & \multicolumn{1}{c}{(deg)} & \multicolumn{1}{c}{(deg)} & \multicolumn{1}{c}{(mag)} & \multicolumn{1}{c}{(mag)} & \multicolumn{1}{c}{(mag)} & \multicolumn{1}{c}{(mag)} & \multicolumn{1}{c}{(mag)} & \multicolumn{1}{c}{(mag)} & \multicolumn{1}{c}{(mag)} & \multicolumn{1}{c}{(mag)} \\

\multicolumn{1}{c}{(1)} & \multicolumn{1}{c}{(2)} & \multicolumn{1}{c}{(3)} & \multicolumn{1}{c}{(4)} & \multicolumn{1}{c}{(5)} & \multicolumn{1}{c}{(6)} & \multicolumn{1}{c}{(7)} & \multicolumn{1}{c}{(8)} & \multicolumn{1}{c}{(9)} & \multicolumn{1}{c}{(10)} & \multicolumn{1}{c}{(11)} \\ \hline

  1  &  87.930710  &  27.056652  &  14.280 $\pm$ 0.004  &  1.195 $\pm$ 0.009  &  1.242 $\pm$ 0.005  &  12.028 $\pm$ 0.021  &  11.585 $\pm$ 0.027  &  11.423 $\pm$ 0.023  &  11.403 $\pm$ 0.044  &  11.309 $\pm$ 0.025  \\
  2  &  87.980881  &  27.032787  &  10.128 $\pm$ 0.025  &  0.483 $\pm$ 0.025  &  0.514 $\pm$ 0.025  &  9.155 $\pm$ 0.021  &  9.110 $\pm$ 0.025  &  8.982 $\pm$ 0.020  &  8.949 $\pm$ 0.034  &  8.934 $\pm$ 0.028  \\
  3  &  87.983437  &  27.020210  &  16.588 $\pm$ 0.011  &  1.577 $\pm$ 0.031  &  1.787 $\pm$ 0.013  &  13.666 $\pm$ 0.019  &  12.969 $\pm$ 0.025  &  12.748 $\pm$ 0.025  &  12.658 $\pm$ 0.036  &  12.671 $\pm$ 0.036  \\
  4  &  87.959221  &  27.028023  &  17.532 $\pm$ 0.019  &  1.244 $\pm$ 0.068  &  1.232 $\pm$ 0.031  &  15.308 $\pm$ 0.036  &  14.780 $\pm$ 0.050  &  14.596 $\pm$ 0.073  &  14.434 $\pm$ 0.042  &  14.448 $\pm$ 0.065  \\
  5  &  87.957176  &  27.040668  &  15.740 $\pm$ 0.005  &  0.918 $\pm$ 0.011  &  1.019 $\pm$ 0.010  &  13.929 $\pm$ 0.023  &  13.515 $\pm$ 0.025  &  13.420 $\pm$ 0.030  &  13.271 $\pm$ 0.043  &  13.291 $\pm$ 0.040  \\

\hline
\end{tabular}
\end{center}
\begin{tablenotes}
\tiny
 {\bf \item Notes:} \\
 \item (1) Serial Number of sources. \\
 \item (2-3) Equatorial coordinates of sources in degrees. \\
 \item (4-6) Photometric magnitudes, colors, and their errors from DFOT \citep{sag12}. \\
 \item (7-9) Photometric catalog either from WIRCam \citep{pug04} or 2MASS PSC \citep{skr06}. \\
 \item (10-11) Photometric catalog from {\it Spitzer} IRAC \citep{faz04}. \\
\end{tablenotes}
\end{table*}

 Fig.~\ref{fig:opt_cmd} shows the distribution of sources detected from the observed optical photometry toward S242 in the $V/(B-V)$ and $V/(V-I)$ color-magnitude (CM) diagrams. The black filled circles represent the sources with photometric uncertainty $<$ 0.1 mag and gray filled circles are those with photometric uncertainty higher than this. The spectroscopically observed stars are numbered and marked with red circles. The blue solid lines represent the locus of the zero-age main sequence (ZAMS) taken from \citet{gir02} and corrected for the cluster distance of 2.08 kpc, and reddening $E(B-V)$ = 0.56 and $E(V-I)$ = 0.70 mag [$E(B-V)/E(V-I) = 0.794$ ; \citealt{coh81}]. In order to shift the ZAMS, we used the extinction $A_{V} = 1.8$ mag (Table~\ref{tab:spectro}) of the main illuminating source BD+26\,980 of the region.
 The photometric catalog of optically observed point sources toward the S242 region is tabulated in Table~\ref{tab:bvi_phot}. A total of 503 sources was detected at least in $B$, $V$, or $I$ band with a limiting magnitude of $V \sim$ 19.4 mag. We obtained a total of 291 sources having counterparts in all the three $B$, $V$, and $I$ bands, within a 18$\arcmin\times$18$\arcmin$ sky area.
 Although the spatial variation of extinction can be non-uniform throughout the region. There may be several causes for the broad distribution of sources in the CM diagrams, such as variable reddening, presence of field stars, binaries, and peculiar stars. However it is difficult to differentiate the cluster members and field stars from these diagrams.

\subsection{Extinction Map Toward the S242 Region}
 \label{ssec:ext_map}

\begin{figure*}
\centering
\includegraphics[width=18.0 cm, height=5.0 cm]{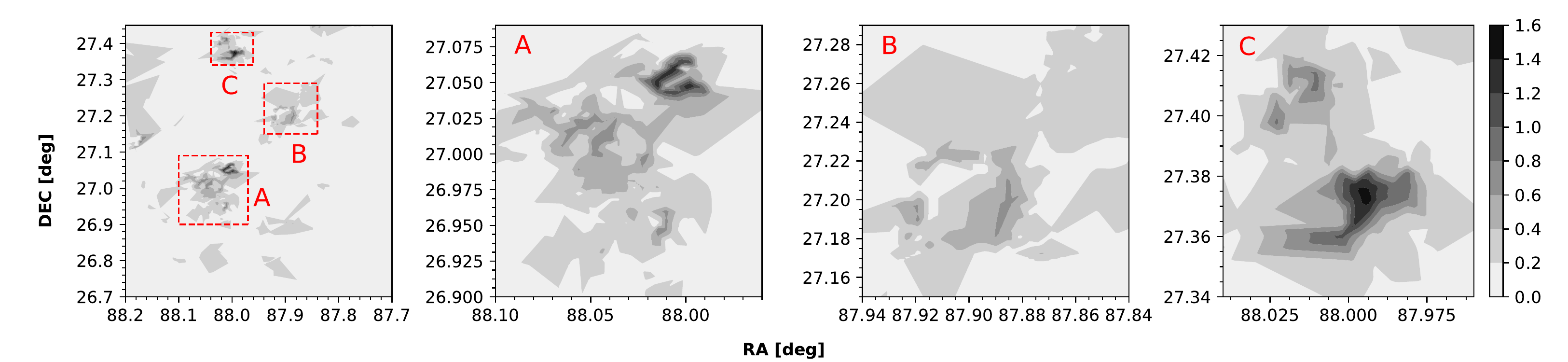}
  \caption{The $K$-band extinction map generated with ($H-K$) colors from the 2MASS catalog. The color bar shows the variation of $A_{K}$ for all the plots. The zoomed-in view of the three subregions (Section~\ref{ssec:ext_map}) is also shown.}
  \label{fig:extinction_map}
\end{figure*}

Discriminating embedded young stars in clusters from field stars is a salient feature in the study of young stellar distributions of interest. Young cluster environments are rich with dust, reducing the density of background stellar contamination \citep{gut05}. Often the nonuniform distribution of dust in these environments makes it far more challenging to get a proper census of detectable background stars \citep{gut05}. In general, the distribution of dust in a cloud can be traced by measurements of the extinction of background starlight produced by the cloud \citep{lad94}.
 In order to map the extinction throughout the S242 region, we used $H$ and $K$ photometry from the 2MASS catalog with photometric uncertainty $<$ 0.1 mag. Since our target area is not completely covered with WIRCam photometry, we used 2MASS catalog to generate the map. The $A_{K}$ values were derived from ($H-K$) colors, following the method outlined in more detail by \citet{gut05}. In brief, we divided the region of our interest into uniform grids of size 15$\arcsec\times$15$\arcsec$. We have taken 20 nearest neighbor sources from the center of each grid to calculate the mean and standard deviation of ($H-K$) color for each grid, excluding the sources whose ($H-K$) values deviated 3$\sigma$ from the mean value. The mean ($H-K$) color for each grid were converted to $A_{K}$, using the reddening law $A_{K} = 1.82 \times [(H-K)_{obs}-(H-K)_{int}]$ from \citet{fla07}. The average intrinsic color $(H-K)_{int}$ of the background population was taken into account to accurately characterize the distribution of extinction toward the region. The $(H-K)_{int}$ color was measured to be 0.2 mag using a nearby unextincted region ($\alpha_{(2000)}$ = 05$^{\rm h}$51$^{\rm m}$54$^{\rm s}$, $\delta_{(2000)}$ = +28$\degr$01$\arcmin$54$\arcsec$) of the sky from 2MASS catalog. The control field was chosen $\sim$ 1$\degr$ away toward the north of S242. In an attempt to reduce the contribution from embedded young stars, only nonexcess infrared sources were used to generate the extinction map. The final extinction map has an angular resolution of 15$\arcsec$ and is sensitive down to $A_V \sim$ 17.8 mag. It is to be noted that the resolution and sensitivity of the map are a function of grid size and number of nearest neighbor sources used. To select the suitable grid size and number of nearest neighbor sources, we performed a series of iterations and found the grid size of $\sim$ 15$\arcsec$ and nearest neighbor source $\sim$ 20 are a good compromise between the sensitivity and signal-to-noise ratio of the extinction map. If the grid size was doubled (30$\arcsec$), the sensitivity dropped to $A_V \sim$ 1 mag.

 The variation of extinction ($A_{K}$) throughout the S242 region is shown in the Fig.~\ref{fig:extinction_map}. The extinction shows a highly nonuniform structure throughout the region. In the map, three distinct and isolated extinction peaks are prominent. They are categorized into three subregions `A', `B' and `C' and their zoomed-in distributions are shown in Fig.~\ref{fig:extinction_map}. The subregion `C' suffers with highest value of extinction $A_{V}$ = 17.2 mag ($A_{K}$ = 1.5 mag); \citep{coh81}, while the subregion `A' has a maximum extinction of $A_{V}$ = 16.1 mag ($A_{K}$ = 1.4 mag). The subregion `B' shows a modest distribution of extinction, with a maximum value of $A_{V}$ = 7.4 mag ($A_{K}$ = 0.7 mag). Among the three extinction complexes, subregions `A' and `C' are the highly extincted regions. These two regions are supposed to be the dominant sites for the next generation star formation. Subregion `B' is supposed to be relatively evolved or its association with molecular clouds is significantly less compared with `A' and `C'. Throughout the region the extinction varies from $A_{V}$ = 1.3--17.2 mag ($A_{K}$ = 0.1--1.5 mag), and the average value of extinction is $A_{V}$ = 3.1 mag ($A_{K}$ = 0.3 mag) with a standard deviation of $A_{V}$ $\sim$ 0.8 mag ($A_{K} \sim$ 0.07 mag). Although, this analysis is limited by sensitivity of the 2MASS survey. The derived extinction can also be underestimated due to the lack of detection of a sufficient number of background stars in the heavily extincted areas.

\subsection{Identification and Classification of YSOs}

 The infrared color-color (CC) space analysis is an efficient tool to identify and characterize the YSOs \citep{lad92}. The YSOs show excess infrared emission due to the presence of circumstellar disks and envelopes and thus occupy certain locations in the infrared CC diagrams. We used IRAC and $JHK$ colors to classify the young sources toward the S242 region. We adopted the IRAC [3.6] and [4.5] $\mu$m along with the WIRCam $H$ and $K$ bands photometry to categorize the pre-main-sequence (PMS) populations using the methods described in \citet{gut09}. Though the infrared excess emission is an essential and powerful membership diagnostic for young and embedded sources, this method also suffers from many potential contaminants. The dominant limitations arise from the extragalactic sources like star-forming galaxies, narrow- and broad-line active galactic nuclei (AGNs), as well as polycyclic aromatic hydrocarbon (PAH) emission excited by young and high-mass stars \citep{gut08}. We used $JHK$ photometry from the WIRCam and the 2MASS catalog as an additional tool for further classification of sources that lack higher wavelength IRAC data. Since a majority of the H$\alpha$ emitters toward \hii regions are considered as classical T Tauri stars (CTTs; \citealt{mey97}), due to the presence of hot and infalling gas accreting from circumstellar disks \citep{bar11}, we used the IPHAS photometry as an additional criteria to detect the young stars showing H$\alpha$ emission.

\subsubsection{Selection of YSOs from IRAC Data}
 \label{sssec:irac_yso}

 Since the S242 region was observed during the {\it Spitzer} Warm Mission \citep{hor12}, the MIR catalog is restricted to [3.6] and [4.5] $\mu$m bands only. Hence, we used IRAC [3.6] and [4.5] $\mu$m along with WIRCam $H$ and $K$ photometry to effectively identify and classify the YSOs. The YSOs in an embedded star cluster suffer high spatially variable extinction due to the presence of natal molecular cloud clumps. We used dereddened IRAC and WIRCam colors by measuring the line-of sight extinction of each source, following the \citet{gut09} classification scheme.

 \begin{figure}
  \includegraphics[width=10 cm,height=7 cm, bb=30 10 370 250]{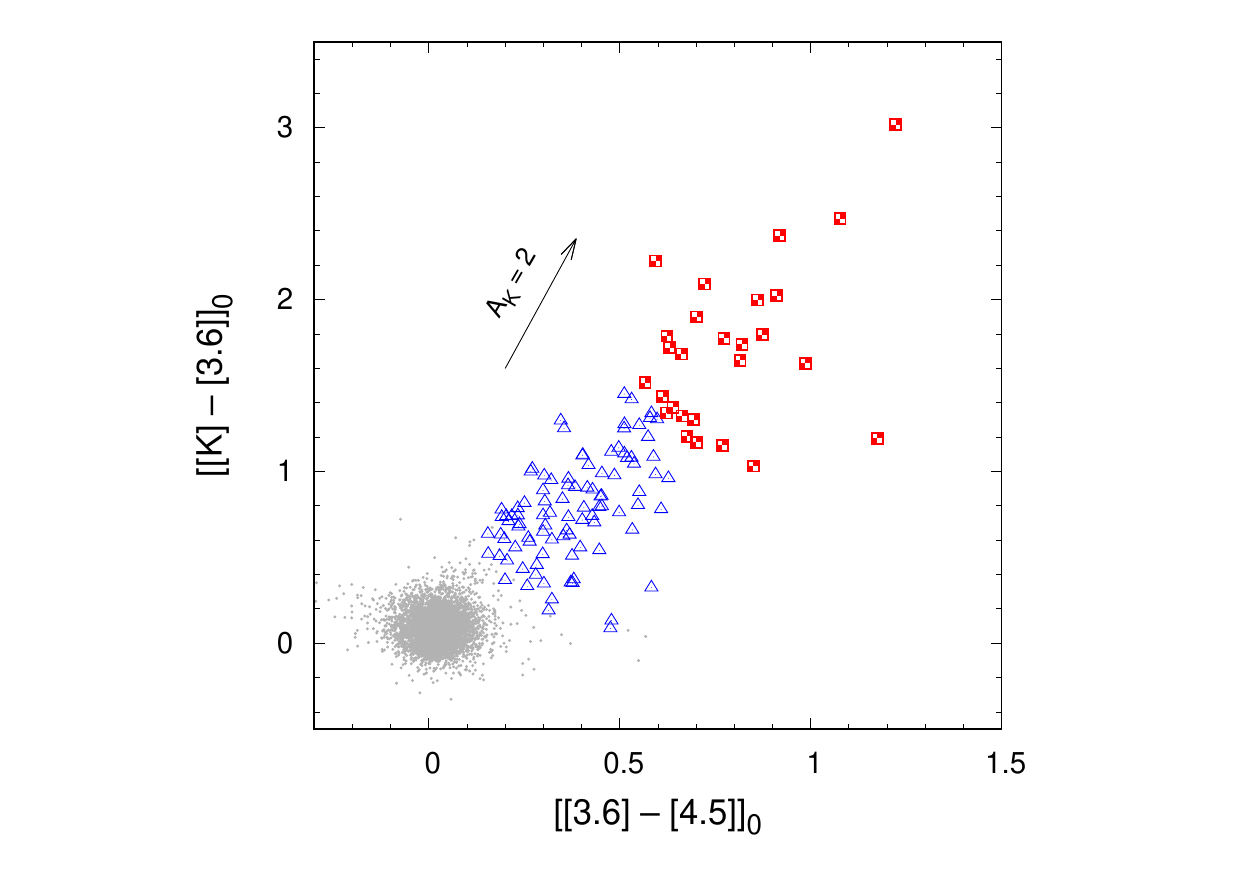}
  \caption{Classification of Class I and Class II sources from the dereddened CC diagram using $H$, $K$, [3.6]~$\mu$m, and [4.5]~$\mu$m photometry, after removing the contaminants. Class I and Class II sources are represented by red squares and blue triangles, respectively. Reddening vector for $A_{K} = 2$ mag is indicated by the black arrow.}
  \label{fig:spitzer_ccd}
 \end{figure}

 Field star contamination was removed by generating the extinction map (Section~\ref{ssec:ext_map}) throughout the region using the 2MASS catalog. To compute the dereddened colors $([[3.6]-[4.5]]_{0}$ and $[[K]-[3.6]]_{0})$ of each source, the extinction of the nearest grid from that source was taken into account. We have identified the YSOs from the dereddened color space using the criteria given by \citet{gut09}, after removing several nonstellar contaminants. An additional brightness cut on the dereddened [3.6] $\mu$m photometry was applied to reduce the inclusion of dim extragalactic contaminants with Class II sources must have $[3.6]_{0} < 14.5$ mag and Class I sources satisfy $[3.6]_{0} < 15$ mag \citep{gut09}. We accomplished a total of 27 Class I and 96 Class II sources for the S242 region using IRAC and WIRCam data. Fig.~\ref{fig:spitzer_ccd} depicts the distribution of Class I (red squares) and Class II (blue triangles) sources in addition to nonexcess field stars (gray dots) in the dereddened CC space. The reddening vector for $A_{K} = 2$ mag is also indicated in the diagram, whereas the color excess ratios were obtained from \citet{fla07}.

 \subsubsection{Additional YSOs from WIRCam and 2MASS Photometry}

\begin{figure}
  \includegraphics[width=10 cm,height=7 cm, bb=30 00 380 250]{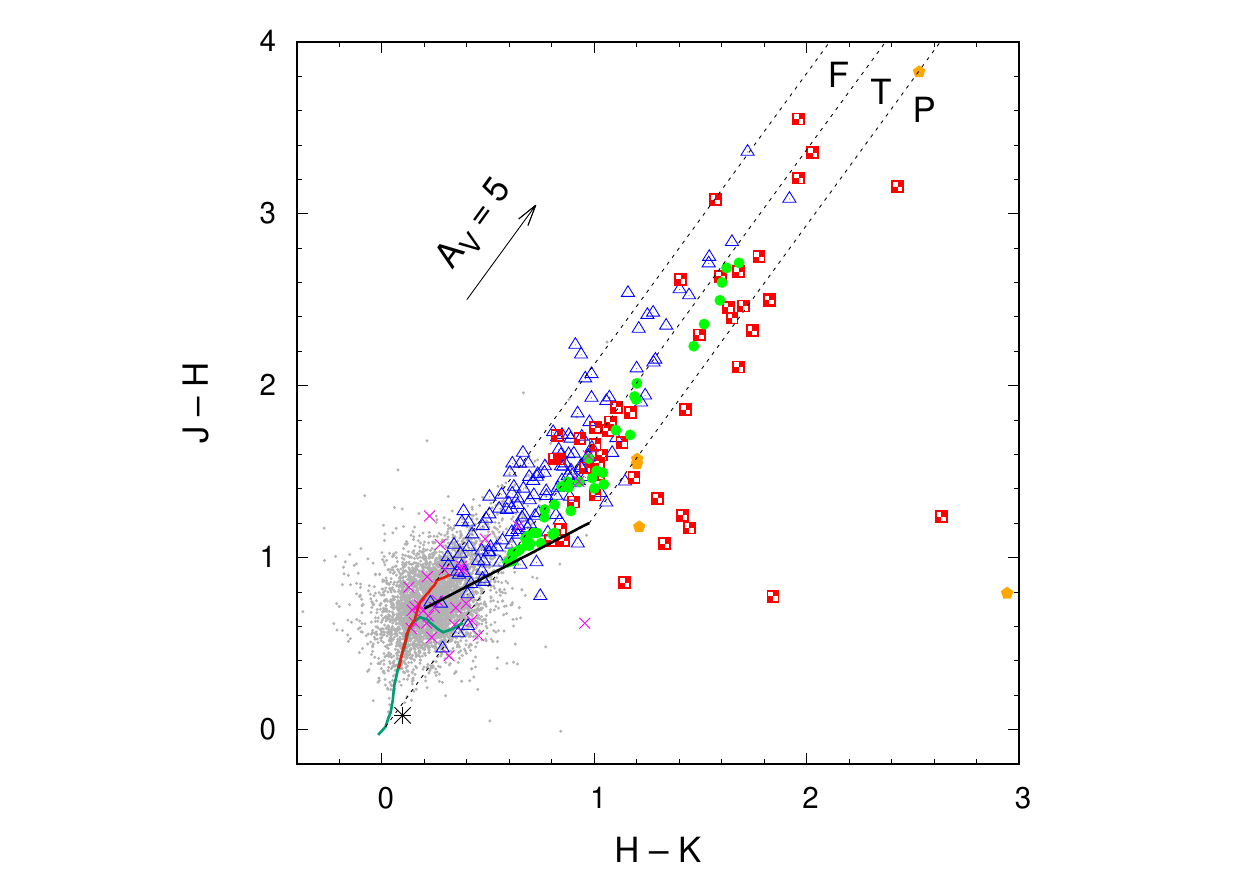}
  \caption{The distribution of YSOs in the NIR CC diagram. The locus of points for unreddened dwarfs and giants are represented by solid green and red lines, respectively. The black solid line indicates the CTTs locus. Red and blue symbols are same as in Fig.~\ref{fig:spitzer_ccd}. The green circle and orange star symbols represent the additional YSOs selected from this diagram. The magenta crosses are the H$\alpha$ emission line sources selected from IPHAS photometry (Section~\ref{ssec:iph}). The main exciting source BD+26\,980 is indicated by a black star symbol and located toward the base of main-sequence locus.}
  \label{fig:wircam_ccd}
 \end{figure}

 We used infrared $JHK$ photometry as an additional selection criteria to distinguish further young objects associated with the molecular dust and cloud. Fig.~\ref{fig:wircam_ccd} shows the distribution of identified YSOs in the $(J-H)/(H-K)$ CC diagram throughout the S242 region. This method is an efficient tool to distinguish the heavily reddened stars having intrinsic infrared excess from those showing normal unreddened photospheric colors \citep{lad92}. The locus of points corresponding to unreddened main-sequence stars and giants is taken from \citet{bes88} and represented by green and red solid lines in Fig.~\ref{fig:wircam_ccd}. The black solid line indicates the CTTs locus, taken from \citet{mey97}. All the photometric magnitudes were converted to the CIT \citep{eli82} system using the relations from \citet{car01}. The three parallel dashed lines show the reddening vectors \citep{lad92}. The reddening laws ($A_{J}/A_{V}=0.265$, $A_{H}/A_{V}=0.155$, and $A_{K}/A_{V}=0.090$) were taken from \citet{coh81}.

 The NIR CC space is divided into three regions, i.e. `F', `T' and `P' \citep{lad92, ojh04, dut15}. The sources located within the `F' region are bounded by the reddening vectors drawn from the unreddened dwarf and giant branches. They have colors similar to normal stellar photospheric sources and are likely to be field stars (main-sequence stars or giants), weak-line T Tauri stars (WTTs), or Class III sources having no or less infrared excess. The `T' region lies within the reddening vectors projected from  main-sequence and CTTs loci. They are considered to be CTTs or Class II sources characterized by the presence of excess infrared emission. The `P' region is located outward to the reddening vector drawn from the points of CTTs locus and the sources show color redder than those in `F'. These sources are most likely Class I objects and show large infrared excess emission due to the heated inner envelopes/atmospheres of the protostars. Although the usefulness of this method to precisely determine the nature of the YSOs is under certain potential limitations \citep{lad92}.

 Using the NIR $JHK$ photometry from WIRCam and 2MASS catalog, we have identified 9 Class I and 59 Class II sources toward S242, with given restrictions on photometric uncertainties $<$ 0.1 mag. Among the 68 YSOs selected from $(J-H)/(H-K)$ CC diagram, 21 sources were previously categorized from IRAC colors (Section~\ref{sssec:irac_yso}). Effectively, we obtained 6 and 41 additional YSOs having colors consistent with Class I and Class II, respectively, using the $JHK$ photometry. The S242 region is enriched with 33 Class I and 137 Class II young objects derived from the combined results of IRAC, WIRCam, and 2MASS photometry. The membership status of the YSOs are further confirmed with the inclusion of distances from the {\it Gaia} DR2 catalog. The distances of the young objects were derived from the available parallax measurements. The photometric and astrometric parameters of the selected YSOs are presented in the Table~\ref{tab:ysos}. While a total of 170 young sources were detected from the infrared photometry, only 28 sources (16.47\%) were supplemented with distances available from the {\it Gaia} DR2. However we do not attempt to further distinguish the Class III objects from WTTs or field population from only this diagram, as this may lead to wrong interpretation \citep{dut15}.

\renewcommand{\tabcolsep}{2.0pt}
\begin{table*}
\begin{center}
\caption{Photometric catalog of YSOs toward S242. The complete catalog is available in the electronic version.}
\tiny
\label{tab:ysos}
\begin{tabular}{|c|c|c|r|r|r|r|r|r|r|c|}
\hline

\multicolumn{1}{|c}{Sl.} & \multicolumn{1}{c}{R.A. (J2000)} & \multicolumn{1}{c}{Decl. (J2000)} & \multicolumn{1}{c}{$g_{P1}$} & \multicolumn{1}{c}{$y_{P1}$} & \multicolumn{1}{c}{$J$} & \multicolumn{1}{c}{$H$} & \multicolumn{1}{c}{$K$} & \multicolumn{1}{c}{[3.6] $\mu$m} & \multicolumn{1}{c}{[4.5] $\mu$m} & \multicolumn{1}{c|}{Distance} \\

\multicolumn{1}{|c}{No.} & \multicolumn{1}{c}{(deg)} & \multicolumn{1}{c}{(deg)} & \multicolumn{1}{c}{(mag)} & \multicolumn{1}{c}{(mag)} & \multicolumn{1}{c}{(mag)} & \multicolumn{1}{c}{(mag)} & \multicolumn{1}{c}{(mag)} & \multicolumn{1}{c}{(mag)} & \multicolumn{1}{c}{(mag)} & \multicolumn{1}{c|}{(pc)} \\

\multicolumn{1}{|c}{(1)} & \multicolumn{1}{c}{(2)} & \multicolumn{1}{c}{(3)} & \multicolumn{1}{c}{(4)} & \multicolumn{1}{c}{(5)} & \multicolumn{1}{c}{(6)} & \multicolumn{1}{c}{(7)} & \multicolumn{1}{c}{(8)} & \multicolumn{1}{c}{(9)} & \multicolumn{1}{c}{(10)} & \multicolumn{1}{c|}{(11)} \\ \hline

\multicolumn{11}{|c|}{Class I sources} \\
\hline

  1  &  88.050873  &  26.980659  &  22.056 $\pm$ 0.250  &  19.045 $\pm$ 0.058  &  17.158 $\pm$ 0.031  &  15.508 $\pm$ 0.024  &  14.647 $\pm$ 0.021  &  12.528 $\pm$ 0.054  &  11.882 $\pm$ 0.040  &  $...$  \\
  2  &  88.073463  &  26.780842  &  19.336 $\pm$ 0.003  &  15.986 $\pm$ 0.004  &  14.647 $\pm$ 0.043  &  17.198 $\pm$ 0.080  &  13.582 $\pm$ 0.044  &  13.373 $\pm$ 0.041  &  13.282 $\pm$ 0.036  &  2203 $\pm$ 607  \\
  3  &  88.010284  &  27.060272  &  $...$   $~~~~~~$  &  $...$   $~~~~~~$  &  16.392 $\pm$ 0.026  &  13.475 $\pm$ 0.016  &  11.624 $\pm$ 0.026  &  9.793 $\pm$ 0.035  &  9.006 $\pm$ 0.038  &  $...$  \\
  4  &  88.039421  &  27.009676  &  $...$   $~~~~~~$  &  $...$   $~~~~~~$  &  18.336 $\pm$ 0.073  &  15.803 $\pm$ 0.024  &  14.082 $\pm$ 0.015  &  12.392 $\pm$ 0.060  &  11.394 $\pm$ 0.036  &  $...$  \\
  5  &  88.038254  &  26.996187  &  $...$   $~~~~~~$  &  $...$   $~~~~~~$  &  18.521 $\pm$ 0.097  &  15.922 $\pm$ 0.023  &  14.220 $\pm$ 0.014  &  12.907 $\pm$ 0.052  &  12.072 $\pm$ 0.049  &  $...$  \\

\hline
\multicolumn{11}{|c|}{Class II sources} \\
\hline

  1  &  88.045395  &  27.021883  &  20.375 $\pm$ 0.012  &  17.074 $\pm$ 0.110  &  14.845 $\pm$ 0.015  &  13.640 $\pm$ 0.014  &  12.956 $\pm$ 0.016  &  12.107 $\pm$ 0.044  &  11.572 $\pm$ 0.038  &  2042 $\pm$ 880  \\
  2  &  87.967293  &  26.984737  &  20.695 $\pm$ 0.053  &  16.880 $\pm$ 0.004  &  15.073 $\pm$ 0.024  &  13.977 $\pm$ 0.025  &  13.394 $\pm$ 0.013  &  12.775 $\pm$ 0.033  &  12.441 $\pm$ 0.035  &  1158 $\pm$ 320  \\
  3  &  87.931961  &  27.002106  &  19.653 $\pm$ 0.016  &  16.094 $\pm$ 0.004  &  14.621 $\pm$ 0.021  &  13.553 $\pm$ 0.025  &  13.005 $\pm$ 0.014  &  12.793 $\pm$ 0.035  &  12.402 $\pm$ 0.030  &  3056 $\pm$ 1351  \\
  4  &  87.992462  &  26.981787  &  21.939 $\pm$ 0.089  &  16.875 $\pm$ 0.016  &  15.129 $\pm$ 0.018  &  13.876 $\pm$ 0.016  &  13.461 $\pm$ 0.032  &  12.808 $\pm$ 0.043  &  12.487 $\pm$ 0.033  &  $...$  \\
  5  &  88.025650  &  27.058596  &  20.403 $\pm$ 0.026  &  15.772 $\pm$ 0.009  &  13.670 $\pm$ 0.017  &  12.012 $\pm$ 0.028  &  10.986 $\pm$ 0.020  &  9.587 $\pm$ 0.035  &  8.965 $\pm$ 0.030  &  $...$  \\

\hline
\end{tabular}
\end{center}
\begin{tablenotes}
\tiny
 {\bf \item Notes:} \\
 \item (1) Serial Number of sources. \\
 \item (2-3) Equatorial coordinates of sources in degrees. \\
 \item (4-5) Photometric magnitudes and their errors from the Pan-STARRS 1 catalog \citep{cha19}. \\
 \item (6-8) Photometric catalog either from WIRCam \citep{pug04} or 2MASS PSC \citep{skr06}. \\
 \item (9-10) Photometric catalog from {\it Spitzer} IRAC \citep{faz04}. \\
 \item (11) Distance of sources from the {\it Gaia} DR2 catalog \citep{gai18}. \\
\end{tablenotes}
\end{table*}

\subsection{H$\alpha$ Emitting Sources from Slitless Spectroscopy and IPHAS Photometry}
 \label{ssec:iph}

\begin{figure}
  \includegraphics[width=10 cm,height=7 cm, bb=30 00 380 250]{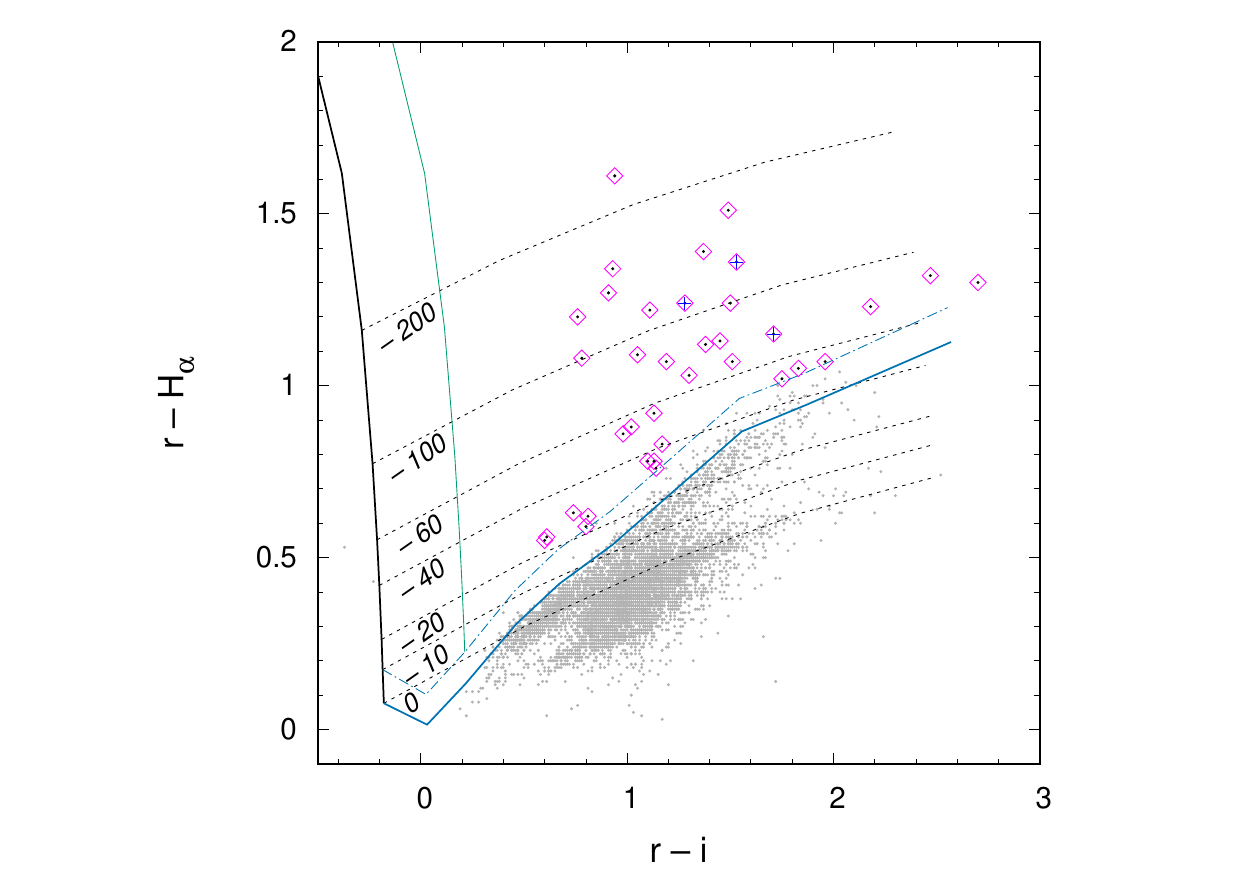}
  \caption{The sources detected in IPHAS photometry are shown in the CC diagram. Blue crosses are the sources detected from slitless spectroscopy and magenta boxes are those selected from IPHAS photometry, with prominent H$\alpha$ emission. Sources with no significant H$\alpha$ emission lines are depicted as gray dots.}
  \label{fig:iphas_phot}
\end{figure}

 Using the slitless spectroscopic data from the HCT, we have visually identified a total of three sources that show counterparts in H$\alpha$ emission. All the three sources are depicted as blue crosses in the CC diagram (Fig.~\ref{fig:iphas_phot}) and were also detected from IPHAS photometry. No extra sources were found from the slitless spectroscopy.

The IPHAS survey provides photometry of fainter emission line objects up to $r$ = 20.5, and $i$ and H$\alpha$ = 19.5 mag with photometric uncertainty limited to 0.1 mag \citep{bar14}. Fig.~\ref{fig:iphas_phot} shows the ($r-i$/$r-$H$\alpha$) CC diagram for the sources detected in the IPHAS catalog toward the S242 region. The solid and dashed blue lines represent the unreddened main-sequence and the expected position of unreddened main-sequence stars with H$\alpha$ emission line strengths of equivalent width (EW) = $-$10 \r {A}. The nearly vertical solid black and green lines show the trend for an unreddened Rayleigh-Jeans continuum and an unreddened optically thick disk continuum \citep{bar14, dut15}, respectively. Whereas the black broken lines are predicted lines of constant net emission EW. The reliable sources are selected by applying the quality criteria of $r$ < 20 mag and photometric uncertainty < 0.1 mag in all three bands in IPHAS DR2. A total of 36 H$\alpha$ emission line stars was selected as lying above the dashed blue line at the level of 3$\sigma$, i.e. the distance between the selected objects and the dashed blue line is larger than three times the average uncertainty in their ($r-$H$\alpha$) color \citep{bar14}. These candidate H$\alpha$ emitters are represented by magenta boxes in the CC diagram. The majority of the H$\alpha$ emitters toward a \hii region are likely to be CTTs \citep{bar11, bar14}. Among the 36 H$\alpha$ emitting objects, 5 sources show infrared counterparts and are classified as Class II sources either from IRAC or $JHK$ colors. The photometric catalog of the H$\alpha$ emitting sources is detailed in Table~\ref{tab:halpha}.

\renewcommand{\tabcolsep}{3.0pt}
\begin{table*}
\begin{center}
\caption{Photometric catalog of H$\alpha$ emitters toward S242. The entire catalog is available in the electronic version.}
\tiny
\label{tab:halpha}
\begin{tabular}{ccccccccccc}
\hline

\multicolumn{1}{c}{Sl.} & \multicolumn{1}{c}{R.A. (J2000)} & \multicolumn{1}{c}{Decl. (J2000)} & \multicolumn{1}{c}{$r$} & \multicolumn{1}{c}{$i$} & \multicolumn{1}{c}{H$\alpha$} & \multicolumn{1}{c}{$J$} & \multicolumn{1}{c}{$H$} & \multicolumn{1}{c}{$K$} & \multicolumn{1}{c}{Distance} & \multicolumn{1}{c}{Class I/} \\

\multicolumn{1}{c}{No.} & \multicolumn{1}{c}{(deg)} & \multicolumn{1}{c}{(deg)} & \multicolumn{1}{c}{(mag)} & \multicolumn{1}{c}{(mag)} & \multicolumn{1}{c}{(mag)} & \multicolumn{1}{c}{(mag)} & \multicolumn{1}{c}{(mag)} & \multicolumn{1}{c}{(mag)} & \multicolumn{1}{c}{(pc)} & \multicolumn{1}{c}{Class II} \\

\multicolumn{1}{c}{(1)} & \multicolumn{1}{c}{(2)} & \multicolumn{1}{c}{(3)} & \multicolumn{1}{c}{(4)} & \multicolumn{1}{c}{(5)} & \multicolumn{1}{c}{(6)} & \multicolumn{1}{c}{(7)} & \multicolumn{1}{c}{(8)} & \multicolumn{1}{c}{(9)} & \multicolumn{1}{c}{(10)} & \multicolumn{1}{c}{(11)} \\ \hline

  1  &  88.006538  &  26.959366  &  18.95 $\pm$ 0.04  &  17.67 $\pm$ 0.02  &  17.71 $\pm$ 0.02  &  17.284 $\pm$ 0.183  &  16.328 $\pm$ 0.185  &  15.913 $\pm$ 0.224  &  1685 $\pm$ 821  &  II  \\
  2  &  87.985306  &  27.042868  &  19.55 $\pm$ 0.05  &  18.79 $\pm$ 0.05  &  18.35 $\pm$ 0.03  &  16.081 $\pm$ 0.069  &  15.362 $\pm$ 0.088  &  14.977 $\pm$ 0.101  &  1221 $\pm$ 1492  &  $...$  \\
  3  &  87.969727  &  27.047367  &  18.71 $\pm$ 0.03  &  17.69 $\pm$ 0.02  &  17.83 $\pm$ 0.02  &  15.342 $\pm$ 0.031  &  14.494 $\pm$ 0.041  &  14.336 $\pm$ 0.056  &  $...$  &  $...$  \\
  4  &  88.241554  &  26.771357  &  19.54 $\pm$ 0.06  &  18.37 $\pm$ 0.03  &  18.71 $\pm$ 0.04  &  16.507 $\pm$ 0.104  &  15.885 $\pm$ 0.122  &  14.876 $\pm$ 999  &  1655 $\pm$ 1312  &  $...$  \\
  5  &  87.995338  &  27.029369  &  18.36 $\pm$ 0.02  &  16.98 $\pm$ 0.01  &  17.24 $\pm$ 0.02  &  16.604 $\pm$ 0.100  &  15.755 $\pm$ 0.116  &  15.670 $\pm$ 0.178  &  735 $\pm$ 86  &  $...$  \\

\hline
\end{tabular}
\end{center}
\begin{tablenotes}
\tiny
 {\bf \item Notes:} \\
 \item (1) Serial Number of sources. \\
 \item (2-3) Equatorial coordinates of sources in degrees. \\
 \item (4-6) Photometric magnitudes and their errors from the IPHAS DR2 source catalog \citep{bar14}. \\
 \item (7-9) Photometric catalog either from WIRCam \citep{pug04} or 2MASS PSC \citep{skr06}. \\
 \item (10) Distance of sources from the {\it Gaia} DR2 catalog \citep{gai18}. \\
 \item (11) Type of class
\end{tablenotes}
\end{table*}

\section{Discussion}
 \label{sec:dis}

\subsection{Spectral Nature of YSOs from NIR CM Diagram}

\begin{figure}
  \includegraphics[width=10 cm,height=7 cm, bb=30 00 380 250]{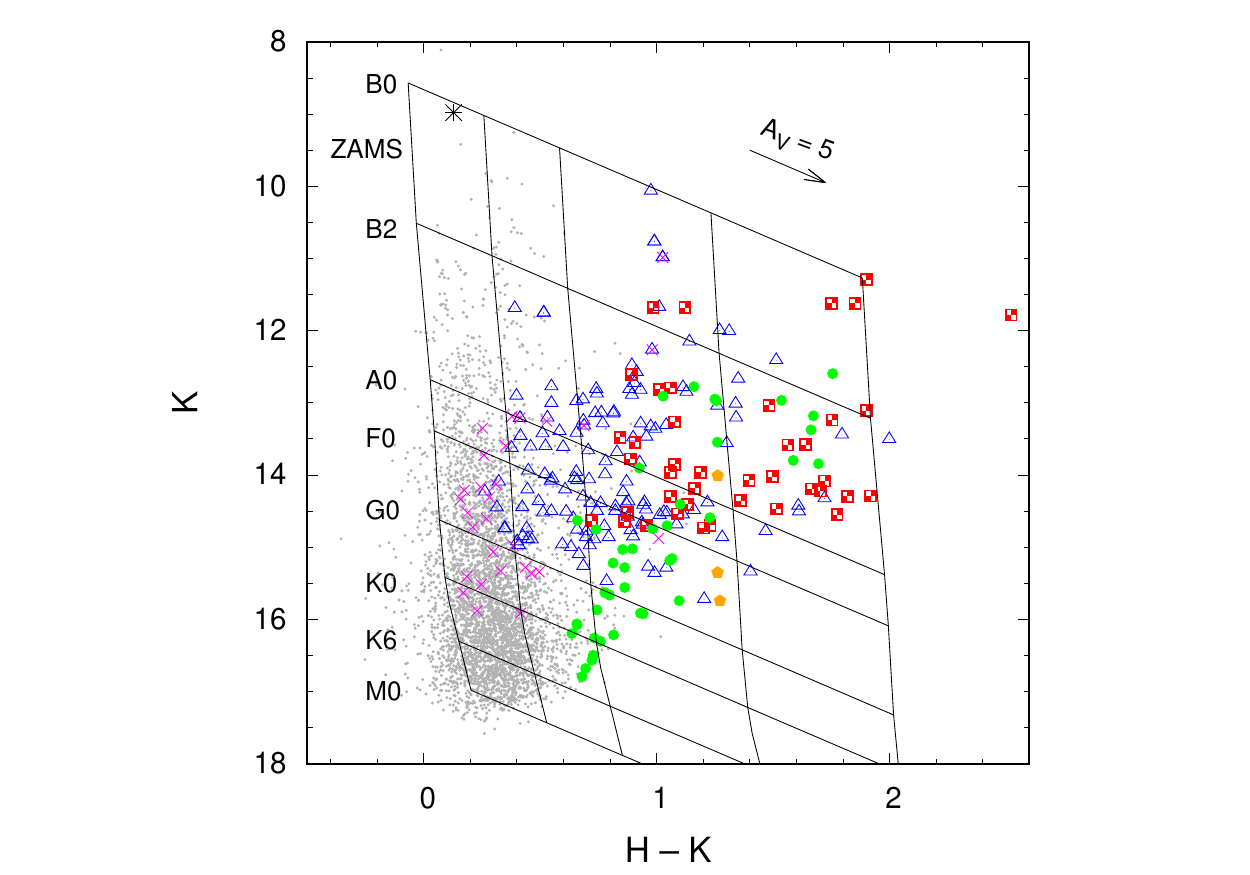}
  \caption{The distribution of YSOs selected from infrared catalogs, H$\alpha$ emitters from IPHAS photometry, and field population toward S242. The nearly vertical lines are the loci of ZAMS \citep{pec13} with visual extinction $A_{V} =$ 0, 5, 10, 20, and 30 mag and shifted for the cluster distance of 2.08 kpc. The slanting horizontal lines represent the reddening vectors corresponding to different spectral types. The symbols used in this figure are the same as in Fig.~\ref{fig:wircam_ccd}.}
  \label{fig:wircam_cmd}
\end{figure}

The NIR CM ($H-K$ versus $K$) space serves as a practical tool to estimate the spectral nature of YSOs by incorporating their distribution in the diagram. Fig.~\ref{fig:wircam_cmd} depicts the YSOs identified from the WIRCam, 2MASS, and IRAC catalogs; H$\alpha$ emitters from the IPHAS photometry, and field population distribution, respectively. The locus of ZAMS \citep{pec13}, reddened by $A_{V} =$ 0, 5, 10, 20, and 30 mag and corrected for the cluster distance of 2.08 kpc (taken from Table~\ref{tab:spectro}) are represented by nearly vertical lines. The slanting horizontal lines are the reddening vectors corresponding to different spectral types. All the symbols in the figure are same as in Fig.~\ref{fig:wircam_ccd}. From the diagram, it is apparent that the YSOs show a large variation in spectral type from K6 to B0, with a majority concentrated within G0 to B2. The main illuminating source of the S242 cluster is located toward the reddening vector corresponding to B0 spectral type, as marked by a black star symbol. Earlier the spectral nature of this star was mentioned as B0\,V, which is quite consistent with our results, and the spectroscopic analysis also reveals it as a B0.5\,V type star. It is prominent that a reasonable amount of YSOs suffer large reddening of about 20 to 30 mag, possibly caused due to the presence of dusty circumstellar envelopes and gaseous environments. So, S242 is a rich stellar cluster, evolving with considerable number of young members and showing wide span in their spectral variation. The gray dots, representing the sources having no or less infrared excess emission, are primarily field population. A majority of them are concentrated within ($H-K$) $<$ 0.6 mag.

\subsection{Estimation of Average Age of the YSOs}
 \label{ssec:yso_age}

\begin{figure}
  \includegraphics[width=10 cm,height=7 cm, bb=40 10 380 250]{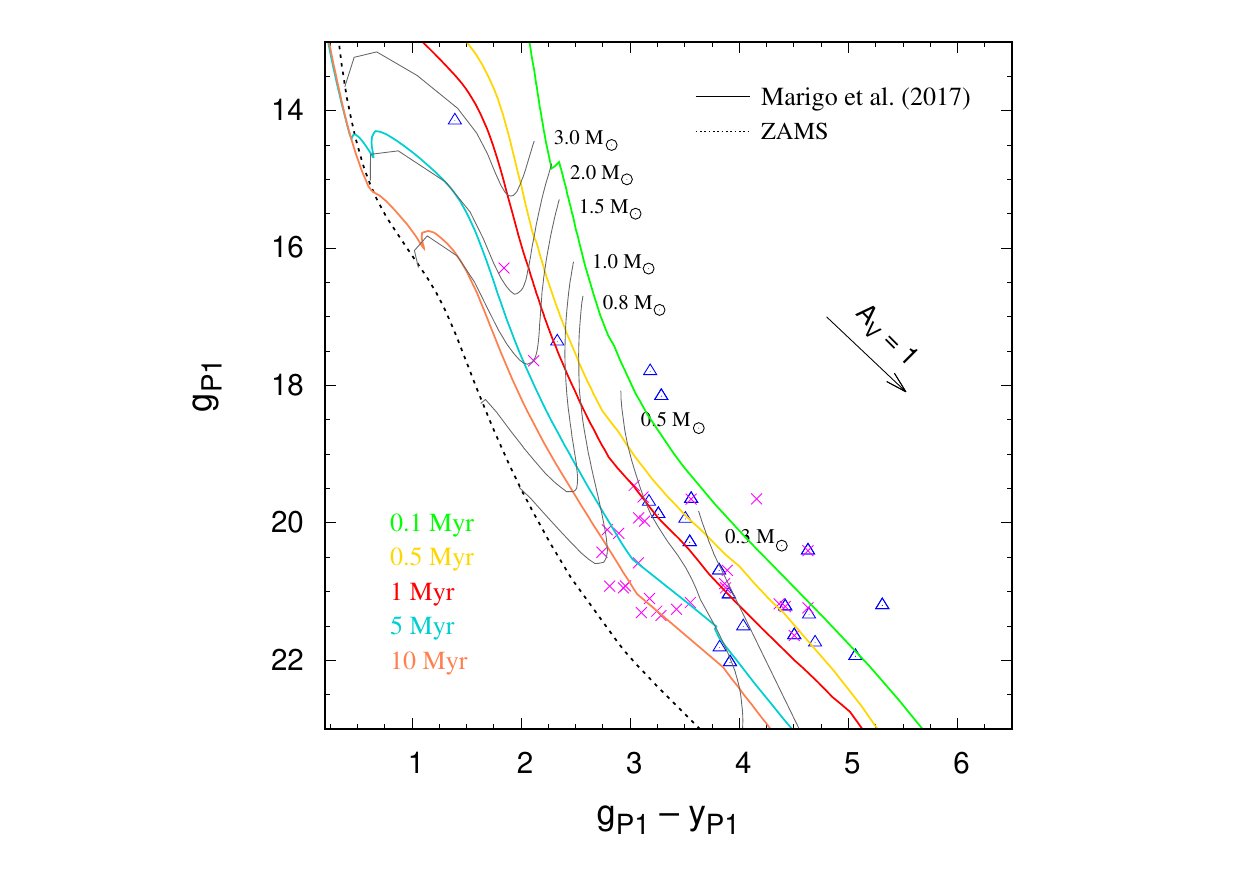}
  \caption{Optical/NIR CM diagram showing the distribution of YSOs and H$\alpha$ emitting sources from the PS1 catalog. The ZAMS and PMS isochrones for ages 0.1, 0.5, 1, 5, and 10 Myr are taken from \citet{mar17}. The evolutionary tracks for different masses are also shown. All the isochrones and tracks are corrected for a distance of 2.08 kpc and reddening $E(B-V)$ = 0.56 mag. The source symbols are the same as in Fig.~\ref{fig:wircam_ccd}.}
  \label{fig:panstarrs_cmd}
\end{figure}

We used Pan-STARRS 1 photometry to diagnose the age spread of the YSOs within the S242 region. Since our observed optical data are not deep enough to detect most of the young sources, we used Pan-STARRS 1 photometry. The magnitude depth of the observed optical photometry is down to $V \sim$ 19.4 mag, whereas Pan-STARRS 1 provides photometry deeper down to {\large g\textsubscript{P1}} $\sim$ 22.5 mag and {\large y\textsubscript{P1}} $\sim$ 20.2 mag for this region. The distribution of YSOs and H$\alpha$ emitting sources in the ({\large g\textsubscript{P1}}$-$ {\large y\textsubscript{P1}}) versus {\large g\textsubscript{P1}} CM diagram is shown in Fig.~\ref{fig:panstarrs_cmd}. The ZAMS and PMS isochrones from \citet{mar17}, corrected for the cluster distance of 2.08 kpc, reddening $E(B-V)$ = 0.56 mag (taken from Table~\ref{tab:spectro}), and $E({\large g\textsubscript{P1}}-{\large y\textsubscript{P1}})$ = 1.26 mag \citep{sch16} are overplotted. The evolutionary tracks for various masses are also shown to characterize the mass spectrum of the YSOs. A notable scatter in the age distribution of the YSO population is observed in Fig.~\ref{fig:panstarrs_cmd}. The ages vary between 0.1 Myr and 10 Myr, with a majority indicating an age around 1 Myr. However, our age determination method can be subject to a few limitations: the use of different PMS evolutionary models can yield different ages \citep{sun00}, the presence of variable extinction, binaries, and variables may also introduce systematic errors \citep{ her94, her99}. We ascertained an average age of the YSOs as 1 Myr. 
Most of the H$\alpha$ emitting sources are relatively evolved compared to the YSOs, as seen from their distribution in the CM diagram. As a larger fraction of the low-mass YSOs lack the reliable photometry, this diagram cannot be used as a suitable tool to estimate mass ranges of the region.

\subsection{Mass Distribution of the YSOs from NIR CM Diagram}
 \label{ssec:yso_mass}

\begin{figure}
  \includegraphics[width=10 cm,height=7 cm, bb=30 00 380 250]{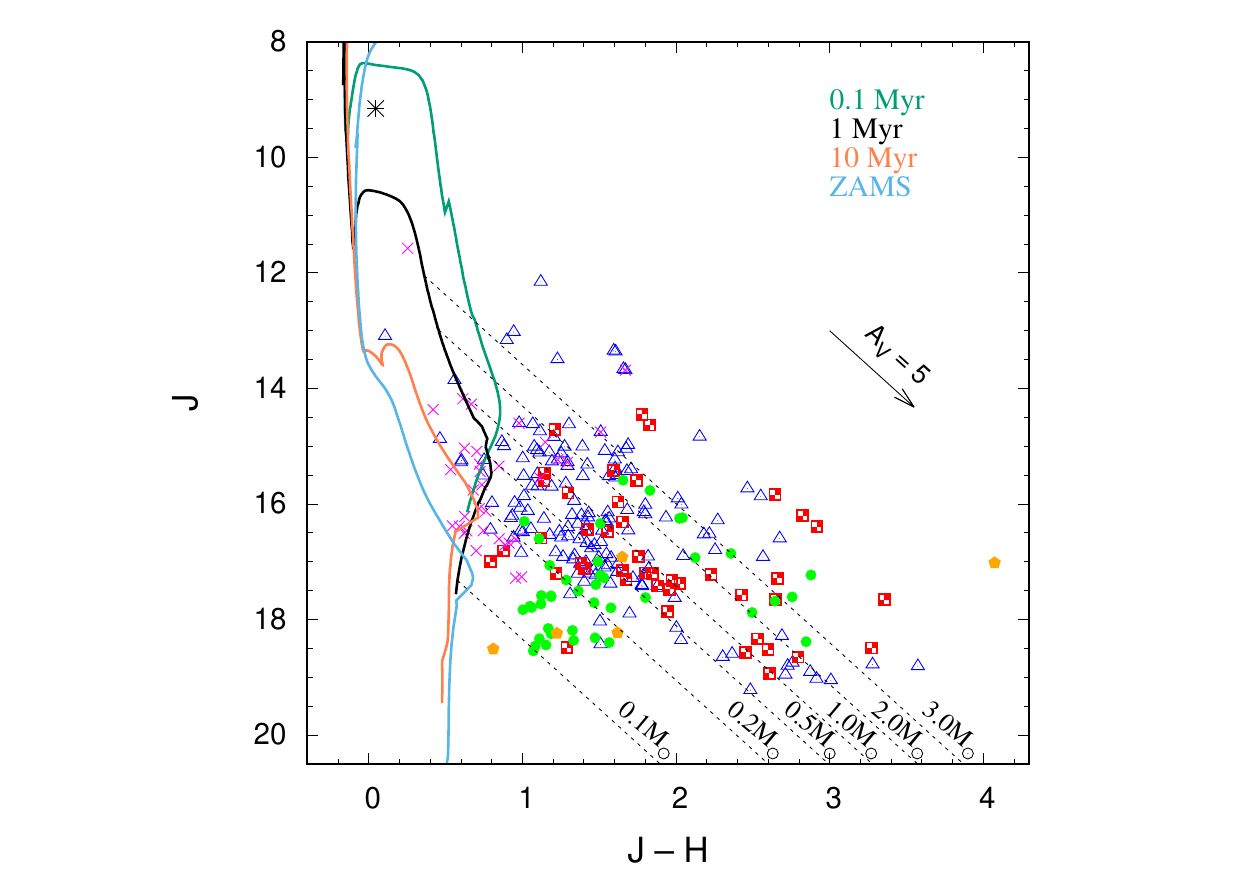}
  \caption{The mass spectrum of the YSOs toward S242 is shown in the NIR CM diagram. The locus of ZAMS \citep{gir02} and PMS isochrones of ages 0.1 Myr, 1 Myr, and 10 Myr \citep{mar17} are indicated by blue, green, black, and red solid lines, respectively. The reddening vectors corresponding to different mass values are represented by black dashed lines. All other symbols are the same as in Fig.~\ref{fig:wircam_ccd}.}
  \label{fig:wircam_cmd2}
\end{figure}

 We used the NIR CM ($J-H$ versus $J$) diagram to estimate the mass ranges of candidate YSOs toward the S242 region. Since the YSOs show excess emission at longer wavelengths, we used ($J-H$) versus $J$ diagram in an attempt to reduce the effect of excess emission \citep{ojh04, ojh11, dut15}. The NIR CM diagrams allow us to manipulate two fundamental parameters of a cluster, the distance and reddening. The change in distance shifts the isochrones vertically, while the parallel slanting lines trace the reddening zones for each extincted mass vector. Masses were estimated by comparing the distribution of young objects with theoretical PMS isochrones in the diagram. In Fig.~\ref{fig:wircam_cmd2}, the blue solid line presents the loci of ZAMS taken from \citet{gir02} and shifted for the cluster distance of 2.08 kpc. The evolutionary models of PMS isochrones of age 0.1, 1, and 10 Myr are taken from \citet{mar17} and indicated by green, black, and red solid lines. The black dashed lines are the reddening vectors corresponding to different mass tracks, respectively. We used the same symbols as in Fig.~\ref{fig:wircam_ccd} to represent the YSOs identified from infrared CC diagrams and H$\alpha$ emitters from IPHAS photometry. We used an average age of 1 Myr (Section~\ref{ssec:yso_age}) to estimate the mass ranges of YSOs. The YSOs show a wide range of variation in their masses with a majority having masses between 0.1 and 3.0 $M_\sun$, as indicated in the figure. It is also observed that the YSOs show larger variation in their color, probably an indication of the combined effect of spatially variable extinction and a weak contribution of excess emission in the $J$ and $H$ bands \citep{ojh11}. Few of the candidate YSOs are seen to be more massive ($>$ 3.0 $M_\sun$) and also located in highly extincted ($A_V$ $\sim$ 10-30 mag) regions. It is to be noted that estimating the stellar masses from the infrared CM diagrams relies on uncertain ages and different distances to the objects. The ambiguity is even more severe among the massive members (early B-type stars). Because the stellar mass can vary significantly when estimated from a 1 Myr PMS isochrone compared to the main-sequence isochrones of a younger age. Therefore, estimating the parameters for the massive members was difficult. Also, there can be several other causes which put constraints on this method, such as the use of separate PMS isochrones, binarity, and variable extinction \citep{hil08}.

\subsection{Spatial Distribution of the PMS Population}
 \label{ssec:spatial}

\begin{figure*}
\centering
\includegraphics[width=11.5 cm,height=10 cm]{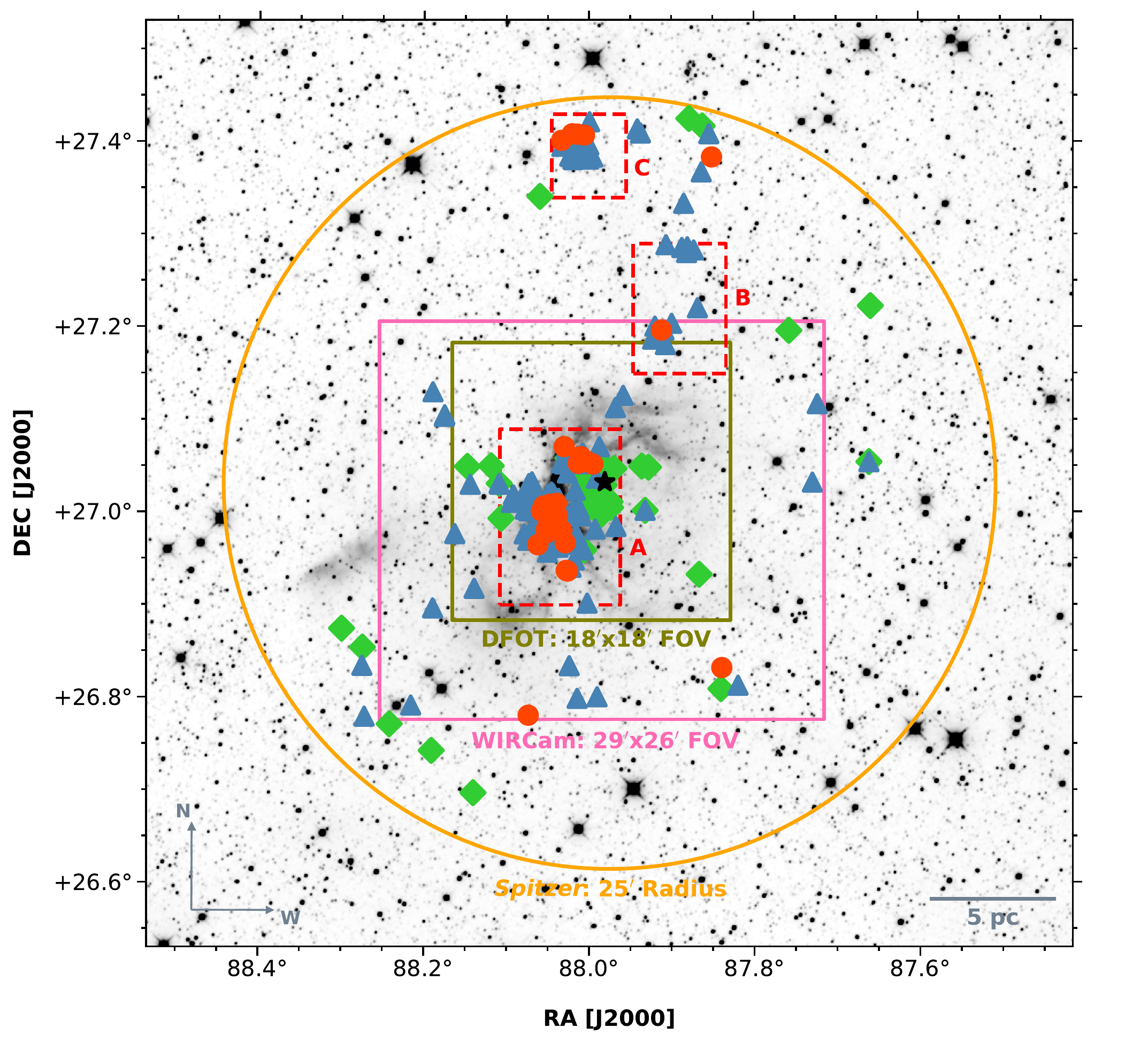}
  \caption{Spatial distribution of the PMS sources, overplotted on the {\it WISE W1} 3.4 $\mu$m mosaic image. The Class I, Class II, and H$\alpha$ emitting sources are represented by red circles, blue triangles, and green boxes, respectively. The main ionizing source BD+26\,980 is represented by a black star symbol. The FOV for different observed and archival catalogs (see Section~\ref{ssec:spatial} for details) are indicated.}
  \label{fig:yso_spatial}
\end{figure*}

The spatial variation of the embedded young stellar population is a footprint of how star formation has progressed throughout space and time in a given region. Fig.~\ref{fig:yso_spatial} manifests the distribution of Class I (red circles), Class II (blue triangles), and H$\alpha$ emission line (green boxes) sources, overlaid on the {\it WISE W1} 3.4 $\mu$m mosaic image. The location of the main ionizing star BD+26\,980 is marked by a black star symbol. The FOV of DFOT optical (18$\arcmin\times$18$\arcmin$), WIRCam NIR (29$\arcmin\times$26$\arcmin$), and other archival photometric catalogs (2MASS, {\it Spitzer}, {\it Gaia} DR2, IPHAS) for a radius of 25$\arcmin$ are indicated in the figure. A noticeable number of the PMS populations are preferentially concentrated around the central core region. A majority of the YSOs are seen to be spatially aligned along an elongated filamentary structure (EFS). From the distribution of young stellar candidates, the spatial extent of the elongated structure is estimated as $\sim$ 43$\arcmin$ (26 pc). Using the $^{13}$CO molecular line data, \citet{dew17} also reported the presence of an EFS of length $\sim$ 25 pc and average width $\sim$ 1.3 pc. A significant number of the Class I sources have formed two discrete groups, one at the central region and another at the northern end of the long scale structure. While Class II type sources also show a similar grouping as Class I sources along with a nice positional coincidence with the long scale feature. The three peak extinction complexes, marked by `A', `B' and `C' (see Section~\ref{ssec:ext_map} for details) are shown by red dashed boxes in the diagram. The prominent clustering of the young populations are nicely match the extinction complexes. The spatial distribution of YSOs along the filamentary-like extinction structure indicates that ongoing star formation is likely occurring toward the cluster.

The number of YSOs in each class is useful to estimate the relative age of the subclusters in a young region \citep{hat07, gut09, sar15}. We used the ratio of number of Class I sources to the number of Class II sources to interpret the evolutionary status of the star-forming region. Since Class I objects represent the earlier phase of star formation compared to Class II sources, the Class I/II ratio serves as a proxy to estimate the relative age of the subclusters \citep{cha08}. The ratio of Class I/II for the subregions `A' and `C' are obtained as 0.31 in both the cases. Whereas, for subregion `B', the ratio is quite low: 0.08. Using a MIR survey of 36 young and active star-forming clusters, \citet{gut09} estimated the median ratio of Class I/II to be 0.27. For the overall S242 region, the ratio is obtained as $\sim$ 0.25, which is in close proximity to the value derived by \citet{gut09}. Studying a nearby star forming complex, \citet{cha08} has shown the ratio of Class I/II to vary between 0.31 and 0.78. \citet{bee10} presented the Class I/II ratio to range from 0.05 to 0.78 for 13 young clusters. \citet{jos13} evaluated the ratio to vary between 0.13 and 0.54 for the different subregions of a large \hii region. The Class I/II ratio is a function of the evolutionary stage of the complex, where the higher ratio indicates a younger cluster \citep{cha08, bee10}. The ratio of Class I/II suggests that the subregions `A' and `C' are the locations of the youngest population for the region and the sources in these groups are evolving almost on a similar timescale. In general, it is predicted from the spatial distribution of the YSOs and the Class I/II ratio that the S242 region is in its early stage of star formation. Comparing the ratio of Class I/II with the earlier reports \citep{cha08, bee10, jos13, sar17}, the average age of the young members in the S242 region is estimated around 1 Myr.
Although, as the subregions `A' and `C' suffer the highest extinction, it may cause undercounting due to the undetectable background stars. Also limitations may occur with distance of the sources and detection sensitivity of the instruments.

\section{Summary and Conclusions}
 \label{sec:sum}

In this work, we report a multiwavelength survey of the stellar content and its characterization toward the S242 region, using combined observed and archival data sets. The key results are summarized as follows:

\begin{enumerate}

\item The slit spectroscopic results confirmed classification of the main ionizing source BD+26\,980 as a massive and early-type (B0.5\,V) star of the region. The spectrophotometric distance of the star is estimated as 2.08 $\pm$ 0.24 kpc, confirming its membership with the region. {\it Gaia} DR2 also provides a similar distance (2.08 $\pm$ 0.19 kpc) for the star. The rest of the spectroscopically observed sources are late-type (either G or K) foreground or background stars.

\item The $K$-band extinction map was generated using a nearest neighborhood technique and the average extinction within the region was estimated as $A_V$ $\sim$ 3.1 mag. From the extinction map (diameter $\sim$ 50$\arcmin$), three distinct extinction peaks were identified toward the region for the first time in this work and their corresponding consequences were discussed.

\item Using the infrared color excess from the combined photometry of NIR (WIRCam and 2MASS) and MIR ({\it Spitzer}) catalogs, a total of 33 Class I and 137 Class II objects were detected within the selected area. Thus the S242 cluster appears as a prominent star-forming region, where a significant number of young stellar populations are found to be forming and evolving together. 

\item The H$\alpha$ emission line objects were detected from IPHAS photometry and slitless spectroscopic observations. We have identified 36 H$\alpha$ emitting sources, which are bonafide young objects toward the S242 region.

\item Using parallaxes from the {\it Gaia} DR2 catalog, the membership status of the classified young population was revealed.

\item Using the Pan-STARRS 1 deep photometry, an average age of the YSOs was estimated as 1 Myr toward the region. From the infrared CM diagram, the masses of the young populations were found to vary between 0.1 and 3.0 $M_\sun$.

\item The cospatial distribution of the young stellar population and the filamentary-like extinction structure is an indication of recent star formation activity within the region. An EFS of length $\sim$ 25 pc is estimated from the projected distribution of the PMS sources. The relatively high fraction ($\sim$ 0.25) of Class I to Class II objects suggests that the YSO population is in a very early stage ($\sim$ 1 Myr) of evolution.

\end{enumerate}

\section*{Acknowledgements}

We thank the anonymous referee for a critical reading of the manuscript and for providing several useful comments and suggestions that significantly improved the scientific content of the paper.
This research work was supported by the S. N. Bose National Centre for Basic Sciences, under the Department of Science and Technology, Government of India. This publication makes use of the observational data from WIRCam, a joint project of CFHT. This work makes use of data products from the 2MASS, which is a joint project of the University of Massachusetts and the Infrared Processing and Analysis Center/California Institute of Technology. This work is based [in part] on observations made with the {\it Spitzer Space Telescope}, which is operated by the Jet Propulsion Laboratory. Archival data are used from Pan-STARRS 1 surveys, operated by the University of Hawaii. This work makes use of the data from IPHAS catalog operated by the Isaac Newton Group of Telescopes. This work presents results from the European Space Agency space mission {\it Gaia}. {\it Gaia} data are being processed by the {\it Gaia} Data Processing and Analysis Consortium (DPAC). We are thankful to the staff of the Indian Astronomical Observatory, Hanle and Centre for Research and Education in Science \& Technology, Hosakote for providing the observational facilities from HCT, operated by the Indian Institute of Astrophysics, Bangalore. The authors are thankful to the staff and members of DFOT, operated by Aryabhatta Research Institute of Observational Sciences, Nainital.

\software{IRAF \citep{tod86, tod93}, DAOFIND \citep{ste92}, DAOPHOT \citep{ste92}, ALLSTAR routine \citep{ste87}, SIMPLE \citep{wan10}}

\end{document}